\documentclass[prl,reprint,doublecolumn,superscriptaddress]{revtex4-1}  

\usepackage{tikz}

\usepackage[sort&compress]{natbib}
\usepackage{soul}
\usepackage{graphicx}
\usepackage{verbatim}
\usepackage{float}
\usepackage{sidecap}
\sidecaptionvpos{figure}{c}
\usepackage{lipsum}
\usepackage{setspace}
\usepackage{dcolumn}
\usepackage{bm}
\usepackage{amssymb}
\usepackage{amsmath}	
\usepackage{mathtools}
\usepackage{color}
\usepackage{multirow}
\usepackage{lipsum}
\usepackage[a4paper]{geometry}
\newgeometry{top=1.25cm,right=1.25cm,left=1.25cm,bottom=1.25cm}
\usepackage[colorlinks=true, urlcolor=blue, pdfborder={0 0 0}]{hyperref}
\hypersetup{
linkcolor=blue,
citecolor=blue
}

\begin{document}

\title{\large{Towards visualisation of central-cell-effects in scanning-tunnelling-microscope images of subsurface dopant qubits in silicon}}

\author{Muhammad Usman} \email{musman@unimelb.edu.au} \affiliation{Centre for Quantum Computation and Communication Technology, School of Physics, The University of Melbourne, Parkville, 3010, VIC, Australia.}

\author{Benoit Voisin} \affiliation{Centre for Quantum Computation and Communication Technology, School of Physics, The University of New South Wales, Sydney, 2052, NSW, Australia.}

\author{Joe Salfi} \affiliation{Centre for Quantum Computation and Communication Technology, School of Physics, The University of New South Wales, Sydney, 2052, NSW, Australia.}

\author{Sven Rogge} \affiliation{Centre for Quantum Computation and Communication Technology, School of Physics, The University of New South Wales, Sydney, 2052, NSW, Australia.}

\author{Lloyd C.L. Hollenberg} \affiliation{Centre for Quantum Computation and Communication Technology, School of Physics, The University of Melbourne, Parkville, 3010, VIC, Australia.}

\begin{abstract}

\noindent
\textbf{\small{Atomic-scale understanding of phosphorous donor wave functions underpins the design and optimisation of silicon based quantum devices. The accuracy of large-scale theoretical methods to compute donor wave functions is dependent on descriptions of central-cell-corrections, which are empirically fitted to match experimental binding energies, or other quantities associated with the global properties of the wave function. Direct approaches to understanding such effects in donor wave functions are of great interest. Here, we apply a comprehensive atomistic theoretical framework to compute scanning tunnelling microscopy (STM) images of subsurface donor wave functions with two central-cell-correction formalisms previously employed in the literature. The comparison between central-cell models based on real-space image features and the Fourier transform profiles indicate that the central-cell effects are visible in the simulated STM images up to ten monolayers below the silicon surface. Our study motivates a future experimental investigation of the central-cell effects via STM imaging technique with potential of fine tuning theoretical models, which could play a vital role in the design of donor-based quantum systems in scalable quantum computer architectures.}}
\end{abstract}
\maketitle

Nuclear or electron spin qubits based on dopant impurities in silicon are promising candidates for the implementation of quantum computing architectures~\cite{Kane_Nature_1998,Hollenberg_PRB_2006,Hill_science_2015,Pica_PRB_2016}. The design and control of robust quantum logic gates demand atomic-scale fabrication~\cite{Fuechsle_NN_2012,Weber_Science_2012} and understanding of the dopant wave functions~\cite{Rahman_PRL_2007,Usman_JPCM_2015, Usman_PRB_2015}. The strength of dopant wave function at the nuclear site is proportional to the contact hyperfine interaction (A), and its long-range spatial distribution underpins the exchange-coupling (J) interaction between dopant pairs -- A and J being two crucial control parameters in the design of quantum logic gates~\cite{Hill_science_2015,Kalra_PRX_2014,Zwanenburg_RMP_2013}. Therefore an accurate description of the dopant wave function has been a long-standing fundamental physics problem~\cite{Kohn_PR_1955, Wilson_PR_1961, Pantelides_Sah_PRB_1974, Martins_PRB_2004, Friesen_PRL_2005}, which has recently attracted renewed research interest~\cite{Rahman_PRL_2007, Pica_PRB_2014, King_1, Wellard_Hollenberg_PRB_2005, Usman_JPCM_2015, Usman_PRB_2015, Salfi_NatMat_2014, Usman_NN_2016}. 

The atomistic description of donor wave function can exhibit central-cell-effects (CCEs) based on underpinning central-cell-corrections (CCCs) implemented in a particular theoretical model~\cite{Usman_JPCM_2015}. These CCEs have been reported to affect the hyperfine Stark shift under the application of electric and strain fields~\cite{Usman_PRB_2015} and the exchange interaction between two P atoms~\cite{Pica2_PRB_2014, Saraiva_arx_2014, Wellard_Hollenberg_PRB_2005}. Currently, benchmarking theoretical descriptions of CCCs has been carried out with respect to measured values of the binding energies and hyperfine with and without electric fields and strain~\cite{Usman_JPCM_2015, Pica2_PRB_2014, Saraiva_arx_2014, Wellard_Hollenberg_PRB_2005, Rahman_PRL_2007}. While such benchmarking of the hyperfine interaction have provided information about the CCCs at the nuclear site, the exchange interaction depends on the long-range spatial distribution of the wave functions, and hence requires investigation of CCEs over a large number of atoms around the P donors. In this work, we demonstrate that scanning tunnelling microscopy (STM) imaging of near surface donors~\cite{Salfi_NatMat_2014,Usman_NN_2016} now provides a way towards much more detailed probing of the spatial distribution of donor wave functions. By theoretically computing STM images based on two different implementations of CCCs employed in the literature, we establish the potential of directly visualising CCEs by STM images, paving the way for high-precision benchmarking of donor states. 

\begin{figure}
\includegraphics[scale=0.28]{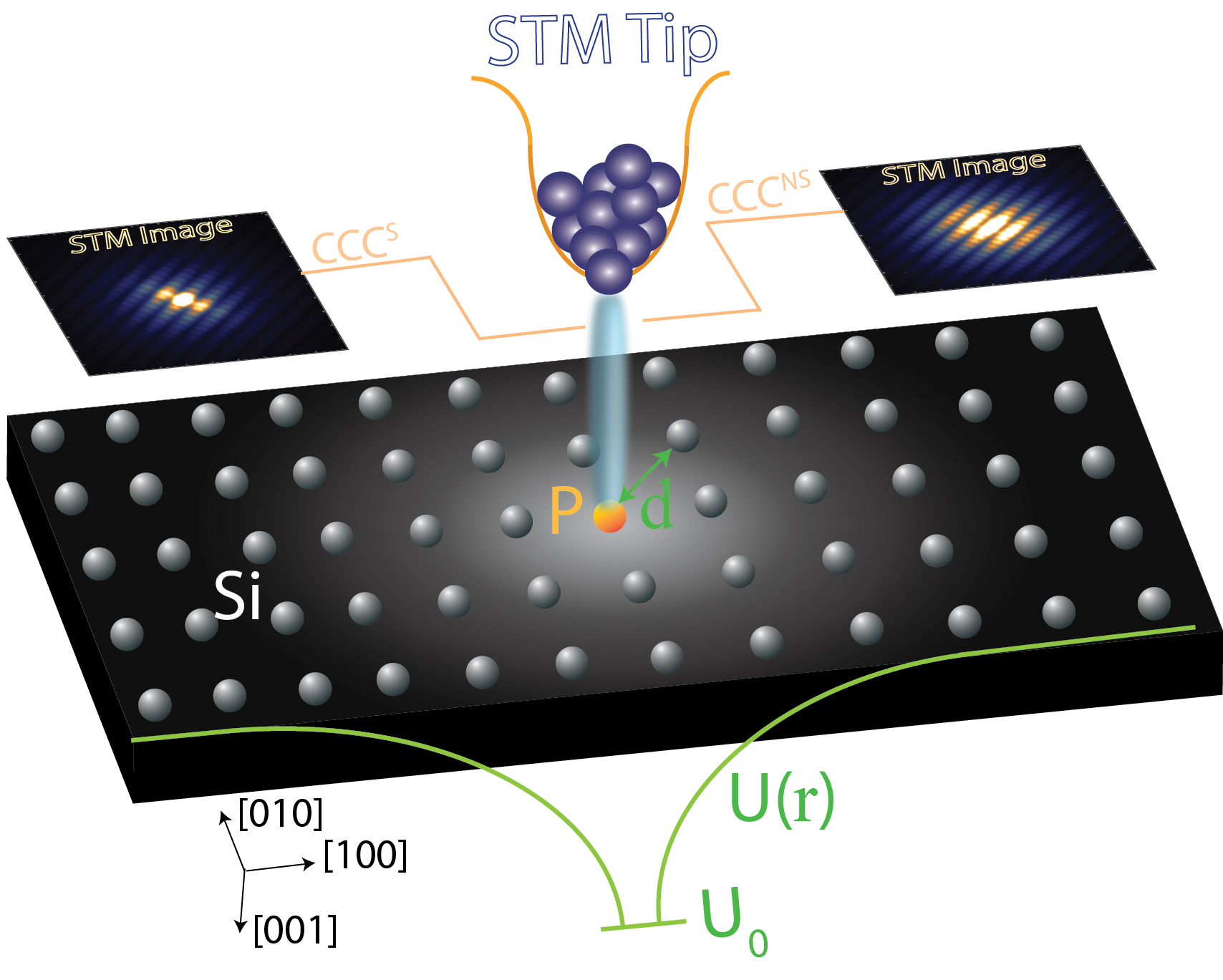}
\caption{\textbf{Schematic illustration of the STM imaging setup:} Schematic diagram of a P donor atom in silicon material, with the background color (light $\rightarrow$ dark) illustrating the potential landscape. The green color plot and labels indicate the following central-cell-corrections (CCCs) to the donor potential: U$_0$ is cut-off potential at the donor site, U($\textbf{r}$) is the long-range tail of donor potential screened by the dielectric constant of silicon, and $d$ is the nearest-neighbour bond-length between P and Si atoms. The calculated tunnelling current image directly depends on the choice of CCCs as illustrated by the two calculated images at the top from the two different implementations of CCCs. }
\label{fig:Fig1}
\end{figure} 

Atomistic approaches to calculating and understanding donor wave functions has a long history. The application of the first principle methods (DFT, ab-initio)~\cite{Overhof_PRL_2004,Smith_arXiv_2016} to the calculation of the dopant wave functions in silicon can capture physics at the donor site, however these approaches are limited to a small number of atoms in the simulation domain due to the requirement of intense computational resources and therefore cannot provide a long-range spatial description of the wave function required for the designing of multi-qubit gates. Momentum based approaches have been engineered to include central-cell and dielectric effects~\cite{Wellard_Hollenberg_PRB_2005}, providing a good description at intermediate scales. For realistic-sized devices with dimensions in the range of 30-50 nm, two commonly employed methods are based on continuum effective-mass~\cite{Koiller_PRB_2002, Pica_PRB_2014, King_1, Pica2_PRB_2014} and atomistic tight-binding~\cite{Usman_NN_2016, Usman_PRB_2015, Ahmed2009, Rahman_PRL_2007} theories, both relying on empirical fitting of the central-cell-corrections (CCCs) to the experimentally measured binding energies and degeneracies of the donor 1s states. This procedure of computing donor states has limited accuracy as it usually relies on benchmarking of the energies without taking into account the spatial distribution of the wave function itself. In this work we employ a comprehensive theoretical treatment of the entire STM-donor system in an atomistic tight-binding framework~\cite{Usman_NN_2016} to investigate whether the impact of central-cell-corrections can be discerned in the context of recently reported STM imaging approach. 
    
Figure~\ref{fig:Fig1} schematically illustrates the setup for STM imaging. We consider a P or As donor located a few monolayers (MLs) below the silicon surface. The wave function distribution of the donor is strongly dependent on central-cell-effects, which modify the donor potential (through a core-cutoff U$_0$, or screening of the potential U($\rm r$) itself) and/or donor-Si bond-length ($d$). STM images are computed from the donor wave functions and are directly related to the underpinning CCEs for up to 10 mono-layers below the hydrogen-passivated (001) silicon surface. As we will show later, a direct theoretical correlation can be established between the computed images and the corresponding donor wave functions. Furthermore, we note that phosphorous (P) and arsenic (As) donor wave functions are primarily different due to the central-cell-effects~\cite{Usman_PRB_2015}, with the CCC induced difference in wave functions being much stronger between the donor species, than between donors of the same species with different CCC implementations. Our results show that the computed STM images for P and As donor wave functions indeed exhibit this stronger effect.
\\ \\
\noindent
\textit{\textbf{\textcolor{blue}{Central-cell-corrections in tight-binding theory:}}} 
\\ \\
\noindent
The implementation details of the central-cell-effects in theoretical models vary considerably in the literature. For instance, the screening of donor potential by silicon dielectric constant has been performed either by a static value~\cite{Martins_PRB_2004, Ahmed2009, Rahman_PRL_2007, Pica2_PRB_2014}, or a non-static $k$-dependent dielectric screening function~\cite{Pantelides_Sah_PRB_1974, Wellard_Hollenberg_PRB_2005, Usman_JPCM_2015, Usman_PRB_2015}. The bond-length deformation (intrinsic strain) around the donor atom has only been taken into account recently~\cite{Overhof_PRL_2004, Usman_JPCM_2015}. To investigate the visualization of the central-cell effects in the computed STM images, here we select the two most often used implementations of central-cell-corrections -- static corrections (CCC$^{\rm S}$) and non-static corrections (CCC$^{\rm NS}$) (\textit{see the supplementary material section S1 for detailed descriptions}). Briefly, the CCC$^{\rm S}$ model is based on a minimal set of central-cell-corrections, where a Coulomb-like donor potential is cut-off at U$_0$=-3.782 eV, which is selected to match the ground state binding energy of the donor atom~\cite{Ahmed2009}. The long-range tail of the potential is screened by a static value of dielectric constant for silicon ($\epsilon \left( 0 \right)$=11.7). The CCC$^{\rm NS}$ model is much more detailed description~\cite{Usman_JPCM_2015}, in which the long-range tail of the potential screened by a non-static dielectric function~\cite{Nara_JPSJ_1965} given by:

\begin{equation}
	\label{eq:Nonstatic_dielectric}
	\frac{1}{\epsilon(k)} = \frac{A^2 k^2}{k^2 + \alpha^2} + \frac{\left( 1-A \right) k^2}{k^2 + \beta^2} + \frac{1}{\epsilon \left( 0 \right) } \frac{\gamma^2}{k^2 + \gamma^2} 
\end{equation}

\noindent where $\epsilon  \left( 0 \right)$ is 10.8, and the fitting parameters A, $\alpha$, $\beta$, and $\gamma$ are 1.175, 0.7572, 0.3223, and 2.044 in atomic units. The cut-off potential (U$_0$) at the donor site is now adjusted to -3.5 eV to match the ground state binding energy. Additionally the nearest-neighbour Si-P bond-lengths are also strained by 1.9\%~\cite{Overhof_PRL_2004}. Both CCC models have been implemented within the atomistic tight-binding framework for the electronic structure calculations~\cite{Usman_JPCM_2015, Rahman_PRL_2007, Klimeck_2, Boykin_PRB_2004}. For the purpose of computing wave function images which are measured via low temperature STM scan~\cite{Salfi_NatMat_2014}, we have not taken free carrier screening into account. Since the P donor based spin qubit devices are typically operated at very low temperatures~\cite{Fuechsle_NN_2012, Weber_Science_2012}, free carrier screening is not important for the presented work. It should be noted that both implementations of CCCs reproduce the ground state binding energy of P atom with high accuracy (with in 1 $\mu$eV of the experimentally measured value of 45.6 meV~\cite{Ramdas_RPP_1981}), however the spatial profiles of the wave functions exhibit clear differences due to the different strengths of donor potential screenings, which is being investigated in this work based on the simulated STM images. It is also important to note that the above model is a very realistic description of the STM experiment. Specifically, although the electric potential applied to the STM tip can in general introduce charges in the substrate, we have chosen the tip voltage to avoid charges during the measurement of the donor state~\cite{Salfi_NatMat_2014}. Conversely, in our experimental setup the tip potential can be chosen to apply an electric field to the quantum state, or even to induce a single-electron quantum dot beneath the tip~\cite{Salfi_arXiv_2017}.

\begin{figure*}
\includegraphics[scale=0.21]{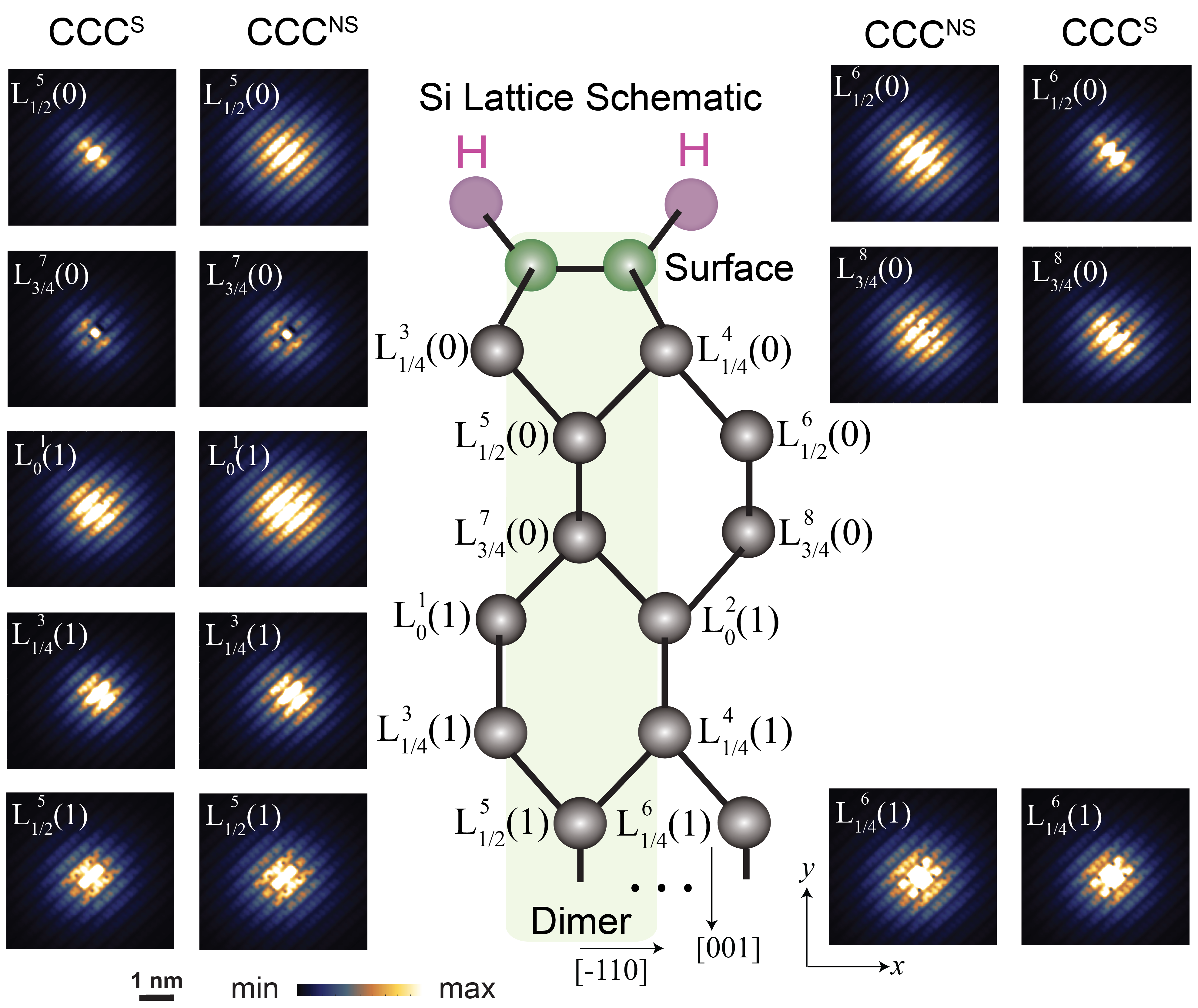}
\caption{\textbf{Calculated real-space STM images:} The calculated real-space STM images are shown for the donor atom positions in the first five unstrained monolayers (ML) below the hydrogen-passivated dimer surface. The labelling of the atom positions follows the notation developed by Usman et al.~\cite{Usman_NN_2016}. Each atom location is labelled as L$_{\rm m}^{\rm i}(\rm n)$, where n=0,1,2,3... defines a plane group, m $\in \{0,1/4,1/2,3/4\}$ is a plane inside n$^{th}$ plane group and i=8m+1 or 8m+2 is an atom within the (m,n) plane. The depth of an atom location from the $z$=0 surface is given by (m+n)$a_0$, where $a_0$ is the silicon lattice constant. The surface atoms in the topmost monolayer are defined as (m,n)=(0,0), with a depth of 0. Note that in the right panel, images for  L$_{\rm 0}^{\rm i}(\rm 1)$ and L$_{\rm 1/4}^{\rm i}(\rm 1)$ are not shown. This is due to the symmetry argument, which leads to the image in right panel being a mere 180$^o$ rotation version of the images of the corresponding positions in the left panel~\cite{Usman_NN_2016}. For each atom position, a direct comparison between the two central-cell-correction models (CCC$^{\rm S}$ and CCC$^{\rm NS}$) clearly highlights their impact on the spatial size and brightness of features in the images.}
\label{fig:Fig2}
\end{figure*}

The important role of central-cell-corrections in the design of donor based quantum devices is evident from the past studies, where based on effective-mass theory~\cite{Saraiva_arx_2014, Pica2_PRB_2014} and band-minimum-basis approach~\cite{Wellard_Hollenberg_PRB_2005}, it has been shown that the strength of exchange interaction between two P atoms is highly dependent on the implementation of donor potential screening in the CCCs. The role of CCCs in the tight-binding theory has recently been shown to accurately reproduce the experimentally measured electric field and strain induced Stark shifts of the hyperfine frequency~\cite{Usman_JPCM_2015, Usman_PRB_2015}. Here the impact of the two implemented CCC models (CCC$^{\rm NS}$ and CCC$^{\rm S}$) on the parameters of interest for the donor based quantum devices is provided in \textit{supplementary figure S1}, which plots the peak amplitudes of wave function charge densities (which is proportional to the contact hyperfine interactions (A)), Z-valley populations, and the exchange-interaction energies ($J$) computed from the two CCC models. The differences between these parameters directly arise due to the atomistic details of the underlying donor wave functions, which considerably vary depending on the selected implementation of the CCC model. It should also be noted that the CCC induced variation in these quantities becomes further pronounced under the application of electric~\cite{Wellard_Hollenberg_PRB_2005} and strain fields~\cite{Usman_PRB_2015} typically employed for controlled operation of quantum devices. 
\\ \\
\noindent
\textit{\textbf{\textcolor{blue}{Calculation of STM images:}}} 
\\ \\
\noindent
The calculation of the STM images (tunnelling current) from the donor wave functions follows the recently reported methodology~\cite{Usman_NN_2016} (\textit{see also details in supplementary section S2}). First the vacuum decay of the donor state at the STM tip position is computed by using the Slater-orbital basis functions~\cite{Slater_PR_1930}. The tunnelling current ($\textrm I_\textrm {T}$) is proportional to the magnitude square of the tunnelling matrix element in accordance with Bardeen's tunnelling theory~\cite{Bardeen_PRL_1961}, whose exact relation with the vacuum-decayed wave function follows the nature of the STM tip orbital configuration as derived by Chen's derivative rule~\cite{Chen_PRB_1990}. To be consistent with the STM measurements reported earlier~\cite{Usman_NN_2016}, we assume the tunnelling current is dominated by $d_{z^2 - \frac{1}{3}r^2}$ orbital state in the STM tip and therefore can be calculated as:   

\begin{equation}\label{func}
{\textrm I_\textrm {T}} (r_0) \varpropto \left\lvert \frac{2}{3}\frac{\partial^2 \Psi_ \textrm D (r)}{\partial z^2} - \frac{1}{3}\frac{\partial^2 \Psi_ \textrm D (r)}{\partial y^2} - \frac{1}{3}\frac{\partial^2 \Psi_ \textrm D (r)}{\partial x^{2} } \right\rvert _{r_0} ^2
\end{equation}

\noindent where $\Psi_{\rm D}$ is the donor wave function and $r_0$ is the position of the STM tip.

\noindent
Here we shall point out that although the $d_{z^2 - \frac{1}{3}r^2}$ orbital might be more localized than s or p type orbitals as discussed in some other work~\cite{Blanco_PSS_2006}, our experimental measurements unambiguously confirm that the dominant orbital in the STM tunnelling current is $d_{z^2 - \frac{1}{3}r^2}$ orbital. This is attributed to the fact that the STM images of subsurface donor atoms in Si have very large spatial extent (overall size of an STM image is 8 nm $\times$ 8 nm), and they exhibit a well-defined symmetry of lattice-sized bright features stemming from valley-related oscillations. We have investigated the role of different tip orbitals and found that the symmetry of the measured STM image features is only reproduced by $d_{z^2 - \frac{1}{3}r^2}$ orbital in the STM tip~\cite{Usman_NN_2016}. This is also consistent with some previous studies where $d_{z^2 - \frac{1}{3}r^2}$ orbital dominance was predicted for transition metal element tips~\cite{Chen_PRB_1990, Chaika_EPL_2010, Teobaldi_PRB_2012}.
\\ \\
\noindent
\textit{\textbf{\textcolor{blue}{Visualisation of CCC effects in STM images:}}} 
\\ \\
\noindent
Figure~\ref{fig:Fig2} plots the real-space STM images computed with the two CCC$^{\rm S}$ and CCC$^{\rm NS}$ models for the available lattice positions within the first five unstrained monolayers below the hydrogen passivated silicon surface. The labelling of the atom positions follows the notation of our earlier work~\cite{Usman_NN_2016}, and has been briefly reiterated in the caption of the figure~\ref{fig:Fig2}. In this study, we do not consider the STM images for the atom positions in the (m,n)=(1/4,0) plane, which is directly connected to the (m,n)=(0,0) surface atoms and therefore experience additional strain due to the formation of surface dimer rows (2$\times$1 reconstruction). The images are normalised and plotted using the same color scale to enable a direct comparison. The images computed from the two CCC models exhibit visible differences, which can be quantified by their spatial extent and color intensity. Furthermore, the line cut profiles through the center of the images along the (-110) direction are also plotted in \textit{supplementary figure S2}, which provide a quantitative comparison between the image features computed from the two CCC models. The size and the brightness of features in STM images are a measure of the spatial distribution of the underlying donor wave function, as the tunnelling current at any given tip position is directly related to the magnitude of the wave function charge density underneath it. The STM images computed from the CCC$^{\rm NS}$ model consistently exhibit brighter features when compared to the CCC$^{\rm S}$ model, thereby implying larger amplitudes of donor wave function density on the surrounding silicon atoms. In order to relate this effect with the CCC potential profile, we also plot in figure~\ref{fig:Fig3} the difference between the donor potentials implemented in the two CCC models. At the donor site, the difference between the two cut-off potentials is $\Delta$U$_0$ is (-3.782) $ - $ (-3.5) = -0.282 eV, indicating that the magnitude of the negative cut-off potential is significantly reduced for the CCC$^{\rm NS}$ model. This leads to a reduction of the wave function density ($|\Psi_D|^2$) at the donor site from 4.05$\times$10$^{30}$ m$^{-3}$ for CCC$^{\rm S}$ to 2.96$\times$10$^{30}$ m$^{-3}$ for CCC$^{\rm NS}$ model, resulting in a better agreement with the measured value~\cite{Feher_PR_1959} of 1.73$\times$10$^{30}$ m$^{-3}$. The spatial profile of the donor wave function on the surrounding silicon atoms is modified by the long-range tail of the potential. Here a larger negative donor potential for the CCC$^{\rm NS}$ model leads to an increase of the donor wave function density on the silicon atoms, resulting in brighter features in the STM images. Therefore, we conclude that for the donor positions in the top-most ten monolayers below the Si surface, the difference between STM images computed from the two CCC models provide a clear evidence of the underlying CCC effects on the corresponding donor wave functions.  

\begin{figure}
\includegraphics[scale=0.28]{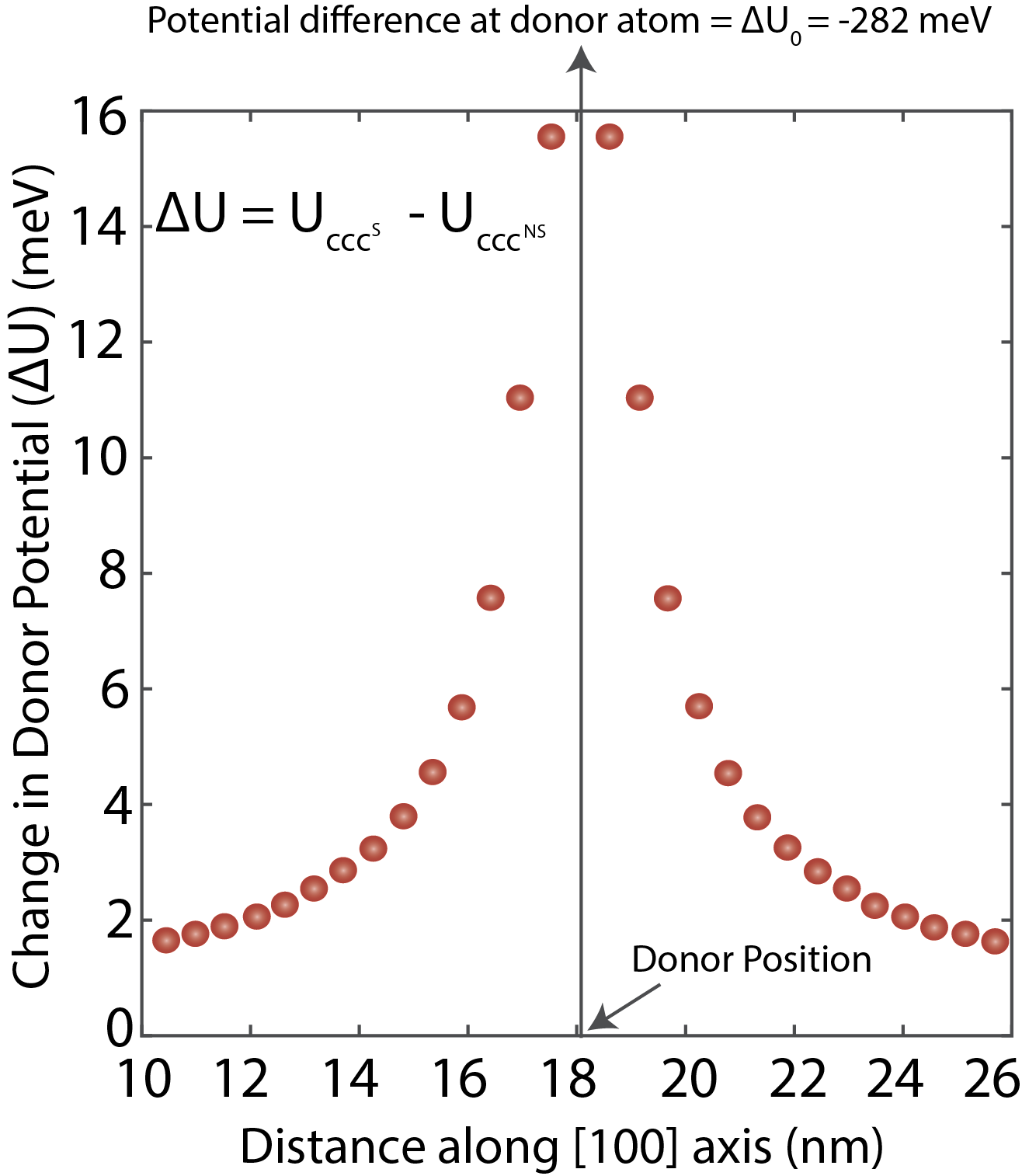}
\caption{\textbf{Difference between CCC donor potentials:} The difference between the donor potentials (U($r$)) for CCC$^{\rm S}$ and CCC$^{\rm NS}$ models is plotted along the [100] direction in the plane of the donor atom. At the donor site, the difference between the cut-off potentials is 282 meV.}
\label{fig:Fig3}
\end{figure} 

\noindent
\\
\noindent
\textit{\textbf{\textcolor{blue}{Quantitative correlation between STM images and wave function parameters:}}} 
\\ \\
\noindent
Next, we quantitatively compare the presence of bright features in the STM images with the actual spatial distribution of the donor wave functions. Here we define the spatial character of an STM image in the donor vicinity in terms of mean radius of its 1/100 contour -- the contour of the normalised image where its peak intensity reduces to 1/100 value. Figures~\ref{fig:Fig4} (a) exhibit exemplary donor images computed from the two CCC models with donor position at L$_{\rm 1/2}^{\rm 5}(\rm 0)$, along with the plots of (1/100)-contours. In figure~\ref{fig:Fig4} (b) we plot the extracted mean radii of the (1/100)-contours as a function of donor depths for both CCC implementations. To represent the dependence of tip-height on the calculations, we plot variation of the results due to a $\pm$ 5\% variation in STM tip height around a fiducial value of 0.2 nm~\cite{Usman_NN_2016}, which indicates that tip height variation contribute only a small variation in the radii of the (1/100)-contours. To correlate this effect with the actual wave function charge densities, we also plot in \textit{supplementary figure S1(c)} the mean radii of (1/100)-contours extracted from the corresponding wave function charge densities in the XY plane of the donor atom. The wave function contour radii for the CCC$^{\rm NS}$ model are larger than for CCC$^{\rm S}$ model, consistently following the trends of STM image contour radii for all of the top-most ten monolayers. This confirms that the extent of the computed STM images accurately represent the spatial distribution profile of the donor wave functions, which is directly modulated by the implementation of the underlying CCC model. 

\begin{figure}
\includegraphics[scale=0.23]{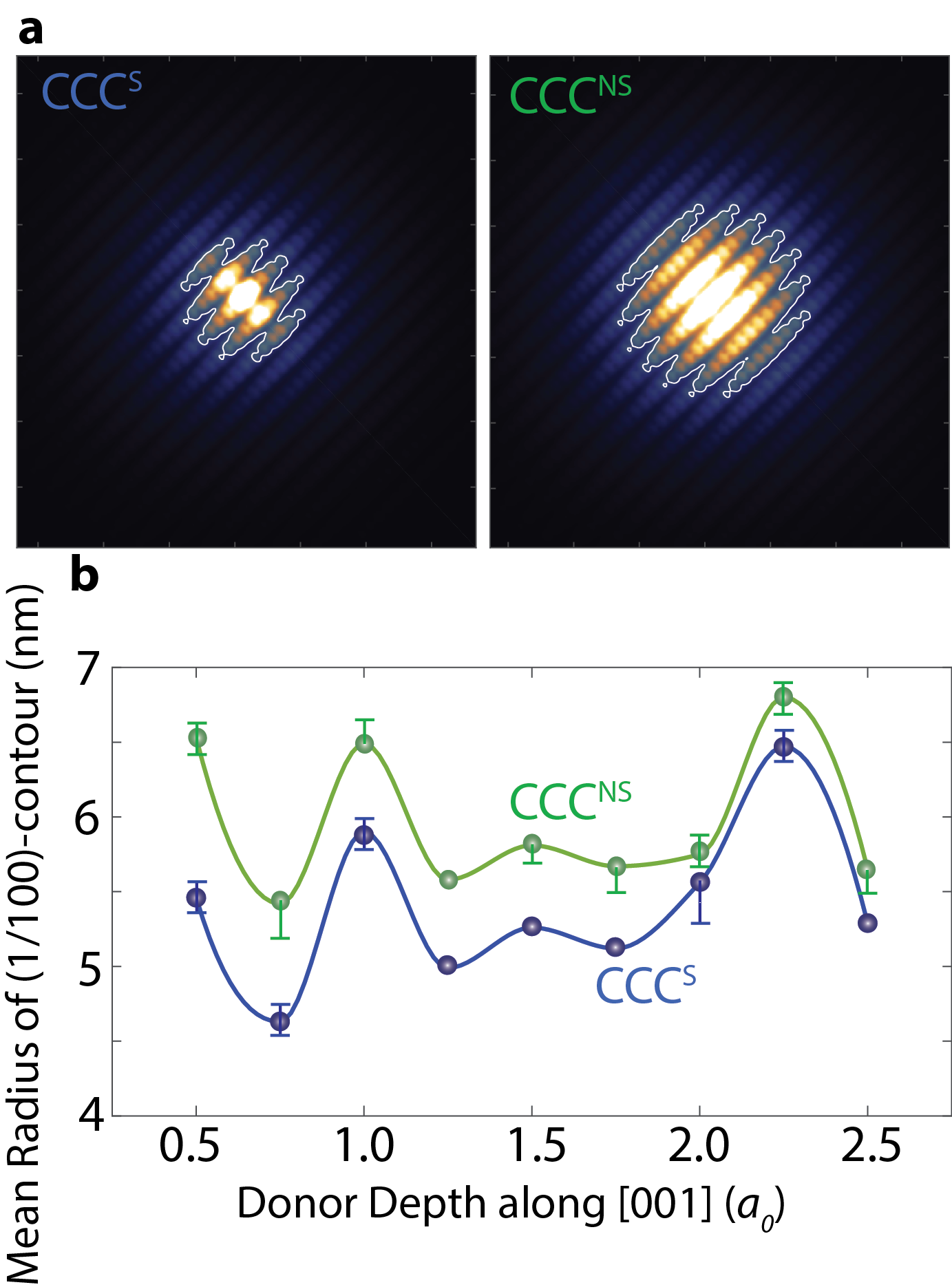}
\caption{\textbf{Correlation between simulated STM images and central-cell correction types:} (a) The (1/100)-contours are plotted on two selected STM images computed with donors at L$_{\rm 1/2}^{\rm 5}(\rm 0)$ positions based on the two CCC models. (b) Mean radii of the (1/100)-contours are plotted as a function of donor depth along the [001] direction. The donor depth is defined in units of $a_0$, which is silicon lattice constant. The data ranges shown indicate contour radii variations corresponding to a $\pm$5\% variation in the STM tip height.}
\label{fig:Fig4}
\end{figure}  

\begin{figure}
\includegraphics[scale=0.38]{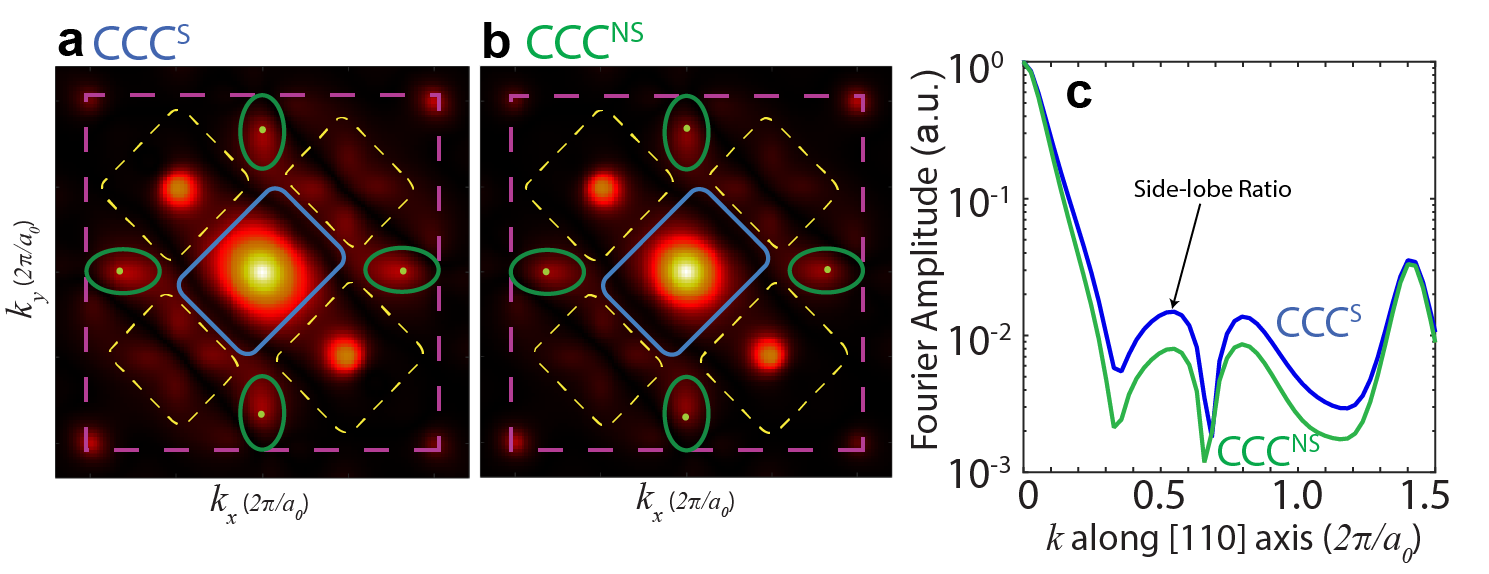}
\caption{\textbf{Fourier transform of STM images:} (a,b) Fourier transform spectra of STM images are plotted for donor location at L$_{\rm 1/2}^{\rm 5}(\rm 2)$ computed from the CCC$^{\rm S}$ and CCC$^{\rm NS}$ models. In each image, the corners of the outer dashed purple box are reciprocal lattice vectors (2$\pi$/$a_0$) $(p,q)$, with $p$=$\pm$1 and $q$=$\pm$1. The ellipsoidal structures corresponding to valleys are found within the green ovals and the green dots indicate the position of the conduction band minima: $k_x$ = 0.82(2$\pi$/$a_0$) ($\pm1,0)$ and $k_y$ = 0.82(2$\pi$/$a_0$) ($0,\pm1)$. The region marked with blue boundary indicates probability envelope and the yellow dashed regions highlight the 2$\times$1 reconstruction-induced features. (c) Line cuts of the Fourier transform spectra from (a) and (b) are plotted along the $k_x$=$k_y$ direction.}
\label{fig:Fig5}
\end{figure}

Whilst the real-space features of the STM images reflect the spatial variations of the donor wave functions, the difference in the spatial extent of the donor wave functions computed with the two CCC models results in a difference in their valley-configurations, which is easier to understand from the Fourier transform of the STM images~\cite{Salfi_NatMat_2014}. The ground state wave function for a bulk donor in silicon is comprised of equal contributions from the six degenerate valleys, however the proximity of the Z interface breaks this symmetry and lifts the degeneracy of valleys, thereby reducing (increasing) the energy (population) of Z-valley, compared to the X- and Y-valleys. The strength of this valley re-population effect is directly related to the interaction of donor wave function with the interface. In our discussion above, we have shown that the wave function charge density for the CCC$^{\rm NS}$ model has larger amplitudes on silicon atoms compared to the CCC$^{\rm S}$ model. Therefore we expect a correspondingly larger interaction of Z interface with the CCC$^{\rm NS}$ state, resulting in a stronger valley re-population. To demonstrate this effect, in figure~\ref{fig:Fig5} (a) and (b) we plot the Fourier transform spectra of computed STM images from the two CCC models for a donor at L$_{\rm 1/2}^{\rm 5}(\rm 2)$ location. If the peak of Fourier spectrum at $k$=0 is normalised to one, the second peak along the $k_x$=$k_y$ line, defined as side-lobe ratio, has been shown to directly reflect the XY valley population of the corresponding donor state~\cite{Salfi_NatMat_2014}. We compare the side-lobe ratios in figure~\ref{fig:Fig5}(c), which indicates a lower amplitude of the slide-lobe-ratio for CCC$^{\rm NS}$ confirming a stronger valley re-population effect (see also \textit{supplementary section S3}), consistent with the expected stronger interface effect. Therefore this work theoretically shows that the CCC induced differences in the valley configurations of the subsurface donor ground states are visible in the Fourier spectra of the corresponding STM images, providing a complementary way to further confirm the validity of the CCC implementation.   

\noindent
\\
\noindent
\textit{\textbf{\textcolor{blue}{Motivation and challenges towards a future experiment:}}} 
\\ \\
\noindent
In this work, we have focused on the effect of CCC on simulated STM images, evidencing differences in the spatial extent and valley population depending on the chosen CCC model. To be able to experimentally investigate CCC mechanism via STM imaging, the first step will require to develop a methodology to identify the donor depth. Depth identification of very shallow donors can be difficult due to the large Z-valley component but has been achieved for subsurface As donors~\cite{Brazdova_PRB_2016}. Furthermore, the fabrication of dopants by atomically-precise fabrication technique~\cite{Weber_Science_2012} is expected to provide a good estimate of the donor positions as recently shown for deeper P donor positions~\cite{Usman_NN_2016}. Although this study is based on the comparison between the computed P atom images, we note that the central-cell related differences are stronger between P and As donor wave functions, which is reflected in the corresponding STM images (\textit{see supplementary section S4}). This could be exploited for the benchmarking of CCCs in the theoretical models. It is also noted that for a future theory-experiment comparison, additional effects such as electric field, strain, and quantum confinement will need to be taken into account, as these influence the STM images and their Fourier transform. Previous tight-binding study of bulk donors in silicon~\cite{Usman_PRB_2015} has included the effects of electric field and strain, which could be extended for the simulation of STM images based on inputs from experimental setup.    

\textit{\textbf{Conclusions:}} In conclusion, this work has compared the computed STM images for two widely used implementations of the central-cell-correction models within the atomistic tight-binding framework and showed that the related effects on Si:P ground state wave functions are visible in both real-space features and frequency profiles of the simulated STM images for donor depths up to ten monolayers below the silicon surface. A quantitative correlation between the parameters such as spatial extent and valley-populations extracted from the computed STM images and the corresponding donor states is established. The theoretical demonstration of the visualisation of central-cell effects in the computed STM images lays the foundation for a future experiment to probe such effects, suggesting a possible route towards high-precision benchmarking of the central-cell-effects in theoretical models. Such a technique could play a vital role in the future studies of donor physics through atomic-scale understanding of the subsurface donor wave functions with higher accuracy compared to the existing empirical approaches.

\textit{\textbf{Conflicts of interest:}} The authors declare no conflicts of interests.

\textit{\textbf{Acknowledgements:}} This work is funded by the ARC Center of Excellence for Quantum Computation and Communication Technology (CE1100001027), and in part by the U.S. Army Research Office (W911NF-08-1-0527). Computational resources are acknowledged from NCN/Nanohub. This work was supported by computational resources provided by the Australian Government through Magnus Pawsey under the National Computational Merit Allocation Scheme.

\newpage
\clearpage

\renewcommand{\thefigure}{S\arabic{figure}}
\setcounter{figure}{0}

\setcounter{equation}{0}

\textbf{\textit{\underline{Supplementary Material Section}}}

\noindent
\\ \\
\textbf{S1. Atomistic Modelling of Donor Wave Functions}
\\ \\
The atomistic simulations of electronic energies and states for a dopant in silicon are performed by solving an \textit{sp$^3$d$^5$s$^*$} tight-binding Hamiltonian. The \textit{sp$^3$d$^5$s$^*$} tight-binding parameters for the Si material are obtained from Boykin \textit{et al}.~\cite{Boykin_PRB_2004}, which have been optimised to accurately reproduce the Si bulk band structure. The phosphorous dopant atom is represented by a central-cell-correction (CCC) model. For this study, we have implemented two central-cell-correction models, namely CCC$^{\rm S}$ and CCC$^{\rm NS}$. In the CCC$^{\rm S}$ model, a Coulomb-like donor potential is screened by static dielectric constant ($\epsilon(0)$) of silicon material and is given by:  

\begin{equation}
	\label{eq:Static_donor_potential}
	U \left( r \right) = \frac{-e^2}{ \epsilon \left( 0 \right) r} 
\end{equation}

\noindent
where $\epsilon(0)$ = 11.9 is the static dielectric constant of Si and $e$ is the charge on electron. The potential is cut-off to U($r_0$)=U$_0$ at the donor site, where the value of U$_0$ is adjusted to match the experimentally measured binding energy spectra of 1s states~\cite{Ahmed2009}.

The second CCC model is much more extensive as it includes intrinsic strain and non-static dielectric screening effects~\cite{Usman_JPCM_2015}. The donor atom is again represented by a Coulomb-like potential, which is  cut-off to U($r_0$)=U$_0$ at the donor site, however it is screened by a $k$-dependent dielectric function and is given by: 

\begin{equation}
	\label{eq:Nonstatic_donor_potential}
	U \left( r \right) = \frac{-e^2}{ \epsilon \left( 0 \right) r} \left( 1 + A \epsilon \left( 0 \right) \mathrm{e}^{- \alpha r} + \left( 1-A \right) \epsilon \left( 0 \right) \mathrm{e}^{- \beta r} - \mathrm{e}^{- \gamma r}  \right)
\end{equation} 

\noindent
where A, $\alpha$, $\beta$, and $\gamma$ are fitting constant and have been numerically fitted as described in the literature~\cite{Nara_JPSJ_1965}. Additionally, the nearest-neighbor bond-lengths of Si:P are strained by 1.9\% in accordance with the recent DFT study~\cite{Overhof_PRL_2004}. The strain dependence is included in the tight-binding Hamiltonian by modifying the interatomic interaction energies in accordance with the Harrison's scaling law~\cite{Boykin_PRB_2004}, which has been previously verified against a number of experimental data sets~\cite{Usman_PRB2_2011, Usman_PRB3_2012, Usman_Nanotechnology_2012, Ahmed2009}.

The size of the simulation domain (Si box around the dopant) is chosen as 40 nm$^3$, consisting of roughly 3 million atoms, with closed boundary conditions in all three spatial dimensions. The effect of Hydrogen passivation on the surface atoms is implemented in accordance with our published recipe~\cite{Lee_PRB_2004}, which shifts the energies of the dangling bonds to avoid any spurious states in the energy range of interest. The multi-million-atom real-space Hamiltonian is solved by a parallel Lanczos algorithm to calculate the single-particle energies and wave functions of the dopant atom. The tight-binding Hamiltonian is implemented within the framework of NEMO-3D~\cite{Klimeck_1, Klimeck_2}.

In the reported STM experiments~\cite{Salfi_NatMat_2014}, the (001) sample surface consists of dimer rows of Si atoms. We have incorporated this effect in our atomistic theory by implementing 2$\times$1 surface reconstruction scheme, in which the surface silicon atoms are displaced in accordance with the published studies~\cite{Craig_SS_1990}. The impact of the surface strain due to the 2$\times$1 reconstruction is included in the tight-binding Hamiltonian by a generalization of the Harrison's scaling law~\cite{Boykin_PRB_2004}, where the inter-atomic interaction energies are modified with the strained bond length $d$ as $(\frac{d_0}{d})^{\eta}$, where $d_0$ is the unperturbed bond-length of Si lattice and $\eta$ is a scaling parameter whose magnitude depends on the type of the interaction being considered and is fitted to obtain hydrostatic deformation potentials. 

\begin{figure*}
\includegraphics[scale=0.3]{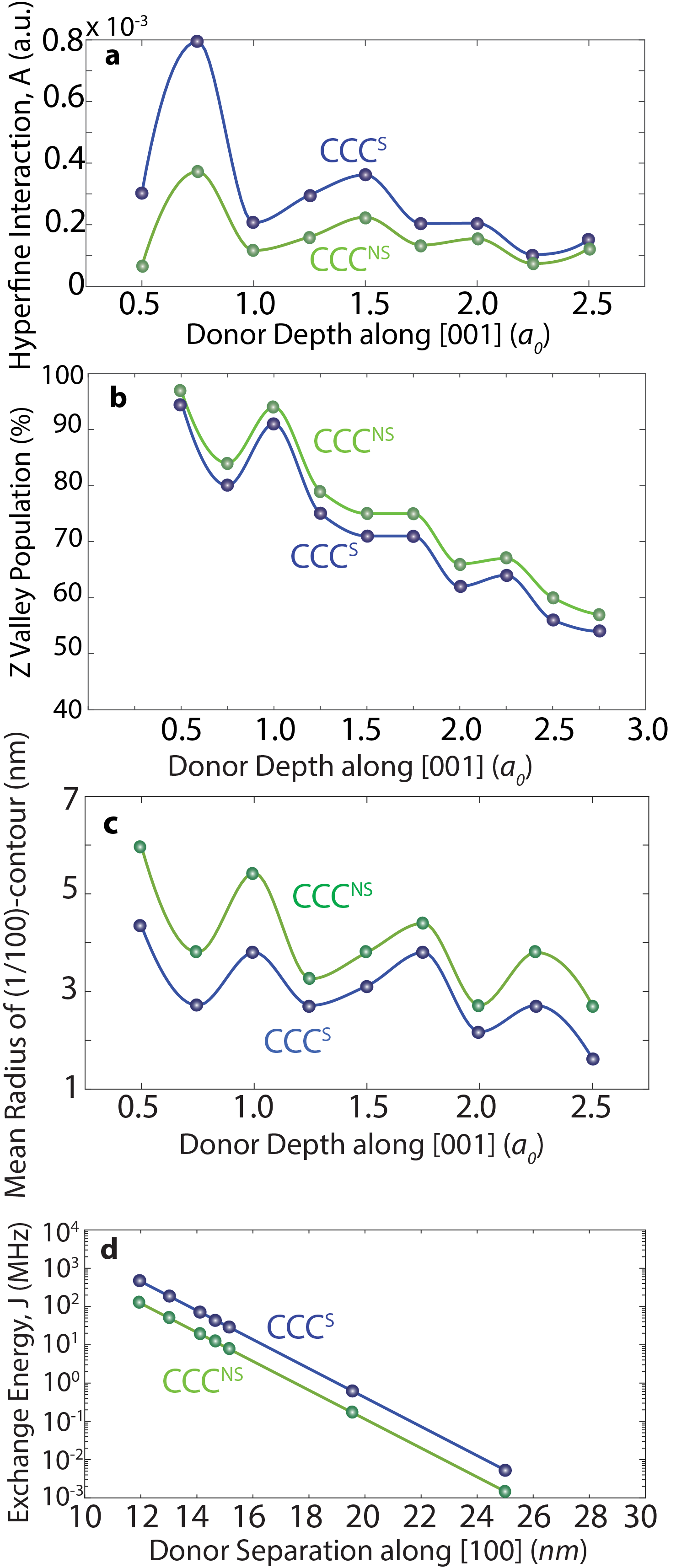}
\caption{(a) The peak amplitude of donor wave function charge density ($|\Psi_ \textrm D|^2$) is plotted as a function of the donor depth below the $z$=0 surface. The contact hyperfine interaction (A) is directly proportional to the peak of $|\Psi_ \textrm D|^2$. (b) The relative Z valley population is plotted as a function of the donor depth below the $z$=0 surface. For a bulk donor ground state, all three valleys equally contribute to the wave function. However the presence of $z$=0 interface lifts the valley degeneracy and increases the weights of Z valleys. (c) Mean radii of the (1/100)-contours extracted from the charge density of the donor wave functions plotted as a function of donor depth along the [001] direction. (d) Exchange interaction is plotted as a function of P donor separations computed from the two central cell models. }
\label{fig:Fig1S}
\end{figure*}

The contact hyperfine interaction (A) is directly proportional to the charge density of the ground state wave function at the donor site $|\Psi_ \textrm D|^2$, and the excited states do not contribute in the magnitude of A~\cite{Usman_PRB_2015}. The Z-valley population of the donor ground states is calculated by following the procedure described in the supplementary information of Salfi et al.~\cite{Salfi_NatMat_2014}. The size of the spatial distribution of the donor wave functions is defined in terms of its mean radius of in-plane (1/100)-contours -- a contour where the amplitude of the normalised wave function decreases to 1/100 value. A direct comparison of these parameters computed from the two CCC models is presented in figure~\ref{fig:Fig1S}. 

One important parameter of interest for exchange-based two qubit quantum logic gate design is the strength of exchange interaction (J) between two P donor atoms. Previous theoretical calculations have shown that the calculation of exchange interaction is very sensitive to the implementation of central-cell corrections~\cite{Wellard_Hollenberg_PRB_2005, Pica2_PRB_2014, Saraiva_1}. We have computed the exchange interaction energies (J) for the two CCC models by using the Heitler-London formalism~\cite{Wellard_PRB_2003} as shown in figure S1 (d), and our calculations indicate a clear dependence of J on the implementation of CCC.   
\\ \\
\noindent
\textbf{S2. Computation of STM Images}
\\ \\
The calculation of the STM images is implemented by coupling the Bardeen's tunnelling theory~\cite{Bardeen_PRL_1961} and Chen's derivative rule~\cite{Chen_PRB_1990} with our tight-binding wave function. In the tunnelling regime, the relationship between the applied bias (V) on the STM tip and the tunnelling current (I) is provided by the Bardeen's formula:

\begin{equation}\label{IV}
{I_T(V)} = \frac{2 \pi e}{\hbar} \sum_{\mu \nu} (1 - f(E_\nu + eV)) |M_ \textrm {DT}|^2 \times \delta(E_\mu - E_\nu -eV)
\end{equation}
\noindent
where $e$ is the electronic charge, $\hbar$ is the reduced Planck's constant, $f$ is the Fermi distribution function, and $M_ \textrm {DT}$ is the tunnelling matrix element between the single electron states of the dopant (denoted by the subscript $\textrm D$) and of the STM tip (denoted by the subscript $\textrm T$). As derived by Chen in Ref.~\onlinecite{Chen_PRB_1990} that the tunnelling matrix element, for all the cases related to STM measurements, can be reduced to a much simpler surface integral solved on a separation surface $\chi$ arbitrarily chosen at a point in-between the sample and STM tip. Therefore,

\begin{equation}\label{tm}
{M_ \textrm {DT}} = \frac{\hbar^2}{2 m_e} \int_{\chi} ( \Psi_ \textrm T^* \nabla \Psi_ \textrm D - \Psi_ \textrm D \nabla \Psi_ \textrm T^* ). d\chi 
\end{equation}
\noindent
where $\Psi_ \textrm D$ is the single electron state of the sample (P or As dopant in Si), $\Psi_ \textrm T$ is the state of the single atom at the apex of the STM tip, and $d\chi$ is an element on the separation surface $\chi$. 

In our calculation of the STM images, we follow Chen's approach~\cite{Chen_PRB_1990}, which reduces equation~\ref{tm} to a very simple derivative rule where the tunnelling matrix element is simply proportional to a functional of the sample wave function computed at the tip location, $r_0$ (for $\chi$ assumed to be at the apex of the STM tip):

\begin{equation}\label{func}
{M_ \textrm {DT}}  \varpropto \Im [\Psi_ \textrm D (r)] 
\end{equation}
\noindent
where the functional of the wave function, $\Im [\Psi_ \textrm D (r)]$, is defined as a derivative (or the sum of derivatives) of the sample wave function -- the direction and the dimensions of the derivatives depend on the orbital composition of the STM tip state. In our recent study~\cite{Usman_NN_2016}, we have shown that the tip orbital that dominates the STM tunnelling current is $d_{z^2 - \frac{1}{3}r^2}$ orbital, for which the equation~\ref{func} becomes:

\begin{equation}\label{func2}
{ M_\textrm {DT}} \varpropto \frac{2}{3}\frac{\partial^2 \Psi_ \textrm D (r)}{\partial z^2} - \frac{1}{3}\frac{\partial^2 \Psi_ \textrm D (r)}{\partial y^2} - \frac{1}{3}\frac{\partial^2 \Psi_ \textrm D (r)}{\partial x^{2} } 
\end{equation}

\noindent
The calculation of tunneling current is based on evaluating equation 6 at the the tip position. For this, we calculate the derivatives of the dopant wave function $\Psi_ \textrm D (r)$ at the tip location, by computing its vacuum decay based on the Slater orbital real-space dependence~\cite{Slater_PR_1930}, which satisfies the vacuum Schr\"{o}dinger equation. Since the derivation of tight-binding Hamiltonian is independent of exact form of basis orbitals, the choice of basis orbitals is arbitrary. However the previous studies~\cite{Lee_PRB_2001, Nielsen_JAP} have shown that the use of Slater-type orbitals works well in the tight-binding theory as they accurately capture the atomic-scale screening of the materials. The analytical form of Slater-type orbitals for silicon material is given in Ref.~\cite{Nielsen_JAP}, which has been used in this work to describe real-space representation of the donor wave function.

It should be noted that the applied bias on STM tip was chosen to induce a small electric field, of the order of $-0.3 \pm 1.9$ MV/m \cite{Salfi_NatMat_2014}. The electric fields of such magnitudes are expected to negligibly perturb the real-space distribution and valley-population of the ground state of subsurface donors. Furthermore, when the STM tip bias was  adjusted to introduce much larger electric field of the magnitude 10 MV/m, the valley population of the donor state was changed by less than 1\%~\cite{Salfi_arXiv_2017}. Therefore in this work, we ignore the effect of electric field induced by the STM tip bias.

Finally, our low temperature STM data precludes the Si-tip model used to explain force-distance spectroscopies. This is not surprising because we work on chemically inert hydrogen terminated surfaces, at low temperature 4.2 K where chemical reactions will be highly suppressed. 
\\ \\
\noindent
\textbf{S3. Fourier Domain Analysis of STM Images}
\\ \\
Figure~\ref{fig:Fig3} plots the Fourier transform of the real-space images of P donor in L$_{\rm 1/2}^{\rm 5}(\rm 0)$, L$_{\rm 1/5}^{\rm 5}(\rm 1)$, and L$_{\rm 1/2}^{\rm 5}(\rm 2)$ locations, computed from both CCC$^{\rm S}$ and CCC$^{\rm NS}$ models. Since the real-space images computed from CCC$^{\rm NS}$ model exhibit larger spatial extent (see Figure 2 in the main text), so the Fourier domain image exhibit lower amplitude in the frequency profile. 

The ground state of a bulk P donor is comprised of equal contributions form six $k$-space degenerate valleys. When the donor atom is brought closer to the Z=0 surface, it increases the Z-valley population and correspondingly X and Y valley populations decrease. The STM images directly probe the donor wave functions, therefore in the Fourier spectrum of the STM images, the $k$-space valley information should be visible. As it turned out, that due to the strong inter-valley interference arising from the cross terms in $|\Psi_\textrm D (k)|^2$ and second derivatives involved due to dominant tunnelling from $d_{z^2 - \frac{1}{3}r^2}$ tip orbital, the components of the Fourier transform of STM images only provide information about the interference of X, Y, and Z valleys. From the previously published analysis~\cite{Salfi_NatMat_2014}, it has been shown that the frequency amplitudes in the so-called side-lobe ratios are qualitatively related to the corresponding X=Y valley populations of the donor wave functions. Figure~\ref{fig:Fig3} (b) plots the Z-valley populations directly extracted from the Fourier spectrum of the donor wave functions, as function of the donor depths, computed from both CCC$^{\rm S}$ and CCC$^{\rm NS}$ models. For both models, the Z-valley population increases as the donor atom becomes closer to the (001) surface. The larger increase in the Z-valley population for the CCC$^{\rm NS}$ model is due to larger spatial extent of the corresponding wave function closer to the donor atom, which strongly interact with the interface as the donor atom is very close to the surface. To correlate this effect with the STM images, we also plot the $k_x$=$k_y$ cut for L$_{\rm 1/2}^{\rm 5}(\rm 2)$ in figure~\ref{fig:Fig3} (c). The amplitude of the frequencies in the side-lobes is weaker for the CCC$^{\rm NS}$ model compared to the CCC$^{\rm S}$ model, which indicates low (high) population for XY (Z) valleys, consistent with the results of figure~\ref{fig:Fig3} (b). As the valley physics is a fundamental parameter of the donor wave functions directly modulated by the underlying CCC implementations, its direct availability in the Fourier spectrum of the STM images provides a viable path towards fine tuning of the CCC parameters directly based on STM measurements.      
\\ \\
\textbf{S4. Comparison of As and P STM Images}
\\ \\
The ground state binding energies of the P and As donors are 45.5 meV and 53.7 meV as measured by the experiment~\cite{Ramdas_RPP_1981}, therefore the As wave function is much more tightly bounded to the donor atom compared to the P wave function. This is captured in the central-cell-effects by having a larger value of the cut-off potential U$_0$ for the As donor, which is 1.3 eV higher than for the P atom~\cite{Usman_PRB_2015}. As a result, the STM images for As donor are expected to be have small spatial size (and larger frequency components). Figure~\ref{fig:Fig4} compares the real space and Fourier space images of the As and P donors in panels (a) and (b), respectively. The donors are placed at the same atomic sites in L$_{\rm 3/4}^{\rm 7}(\rm 0)$, L$_{\rm 3/4}^{\rm 7}(\rm 1)$, and L$_{\rm 3/4}^{\rm 7}(\rm 2)$. The difference between the STM images of As and P donors are quite evident, which indicates that STM imaging technique could differentiate between As and P donors. In fact, in future, if an experiment is performed by placing P and As atoms at the same lattice location, a direct comparison between the two measurements could provide a way to fine tune central-cell effects.   

\newpage

\begin{figure*}
\includegraphics[scale=0.21]{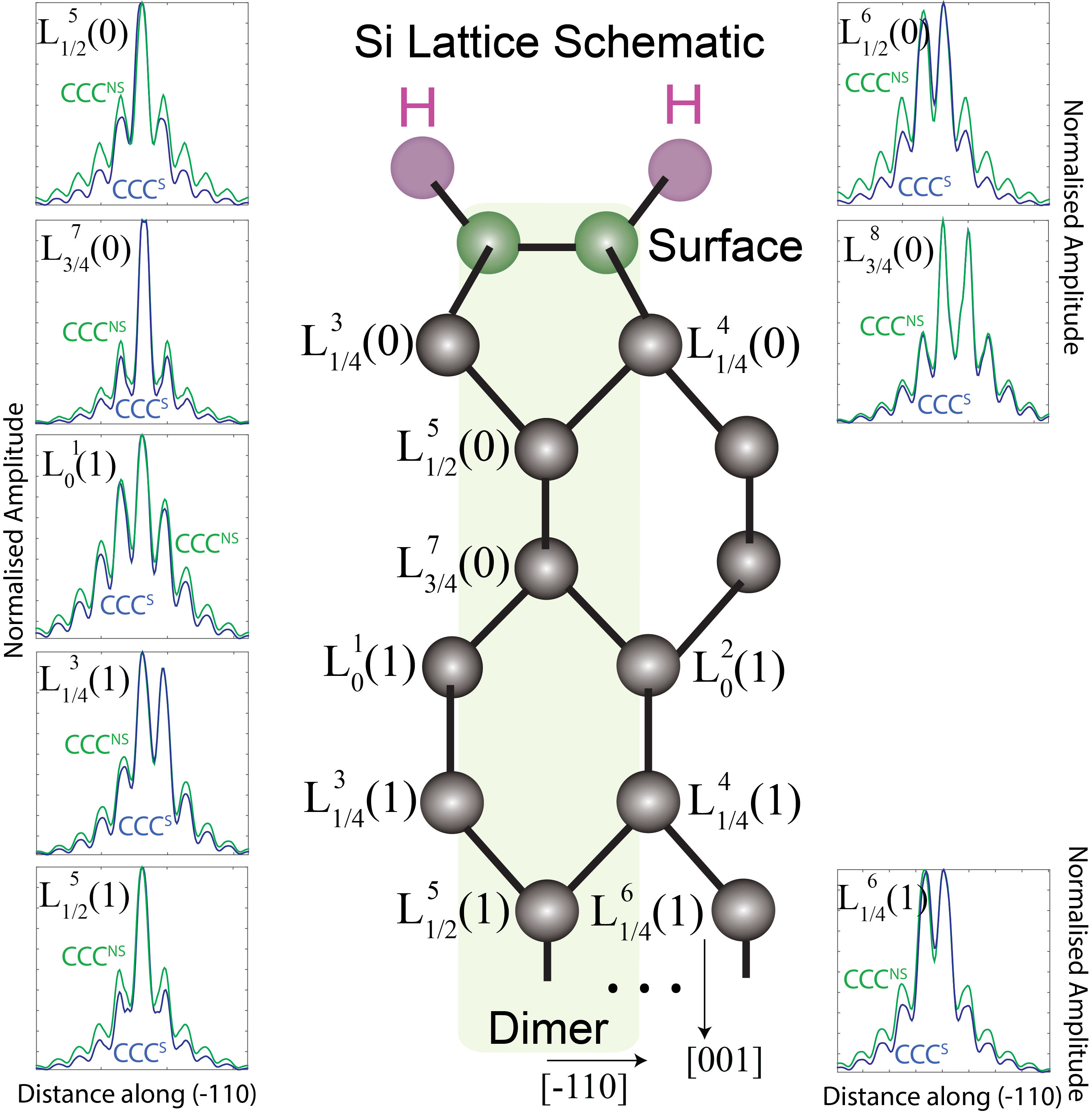}
\caption{The calculated real-space STM images are shown in the main text figure 2 for the donor atom positions in the first five unstrained monolayers (ML) below the hydrogen-passivated dimer surface. Here we plot the line cut profiles through the center of images along the (-110) direction, quantitatively highlighting the difference between the STM images computed from the two CCC models.}
\label{fig:Fig2}
\end{figure*}

\begin{figure*}
\includegraphics[scale=0.25]{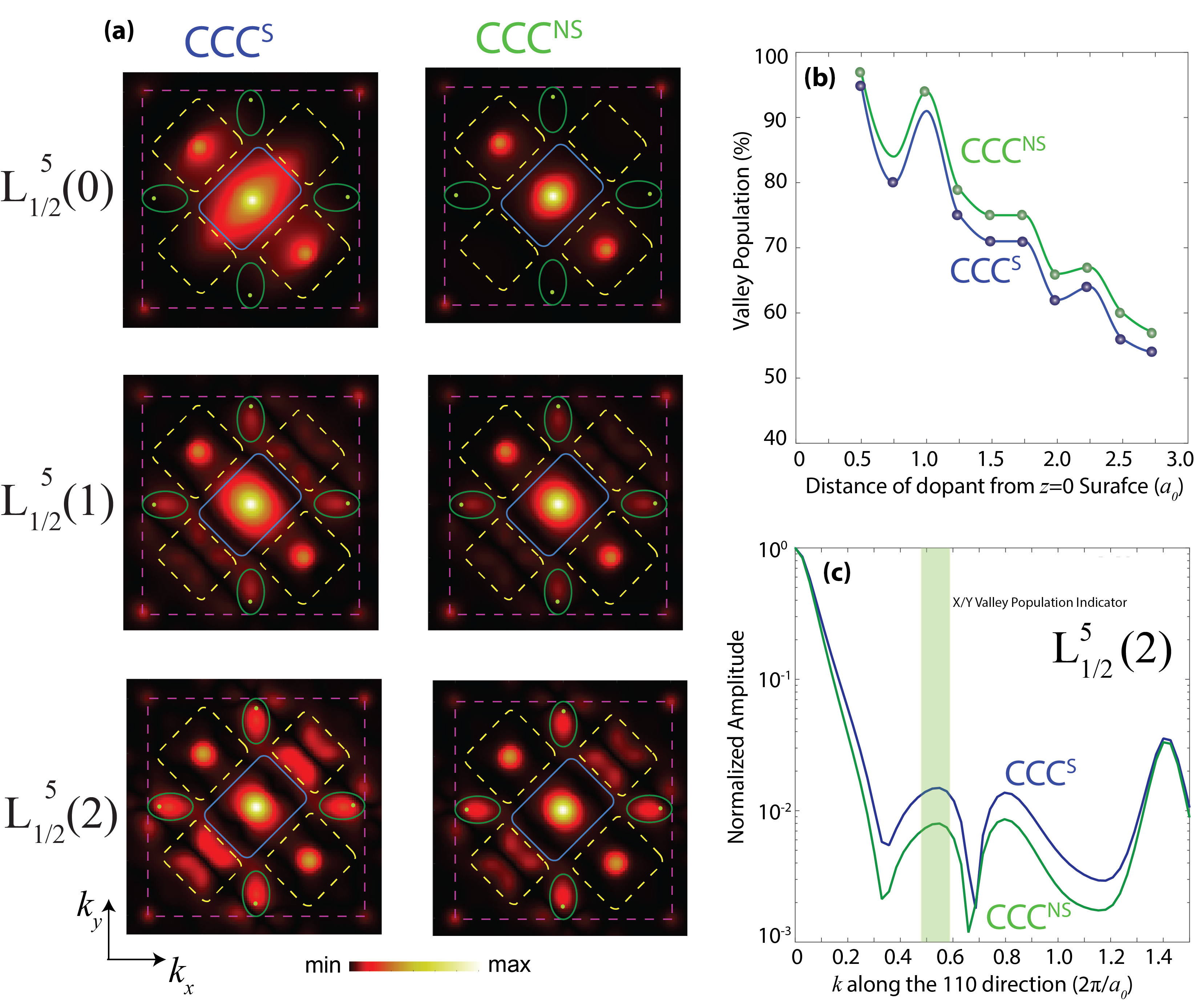}
\caption{(a) Fourier transform spectrum of P donor images in L$_{\rm 1/2}^{\rm 5}(\rm 0)$, L$_{\rm 1/5}^{\rm 5}(\rm 1)$, and L$_{\rm 1/2}^{\rm 5}(\rm 2)$ locations computed from the CCC$^{\rm S}$ and CCC$^{\rm NS}$ models. In each image, the corners of the outer dashed purple box are reciprocal lattice vectors (2$\pi$/$a_0$) $(p,q)$, with $p$=$\pm$1 and $q$=$\pm$1. The ellipsoidal structures corresponding to valleys are found within the green ovals and the green dots indicate the position of the conduction band minima: $k_x$ = 0.85(2$\pi$/$a_0$) ($\pm1,0)$ and $k_y$ = 0.85(2$\pi$/$a_0$) ($0,\pm1)$. The region marked with blue boundary indicates probability envelope and the yellow dashed regions highlight the 2$\times$1 reconstruction-induced features. (b) Z-valley population of the P donor ground states, directly extracted from the Fourier transform of the donor wave function computed from CCC$^{\rm S}$ and CCC$^{\rm NS}$ models. (c) Line cuts of the Fourier transform spectra of the STM images for P donors in the L$_{\rm 1/2}^{\rm 5}(\rm 2)$ position, along the $k_x$=$k_y$ direction. The amplitudes of Fourier transform in the shaded region are directly related to the X=Y valley population of the corresponding donor wave function.}
\label{fig:Fig3}
\end{figure*}

\begin{figure*}
\includegraphics[scale=0.2]{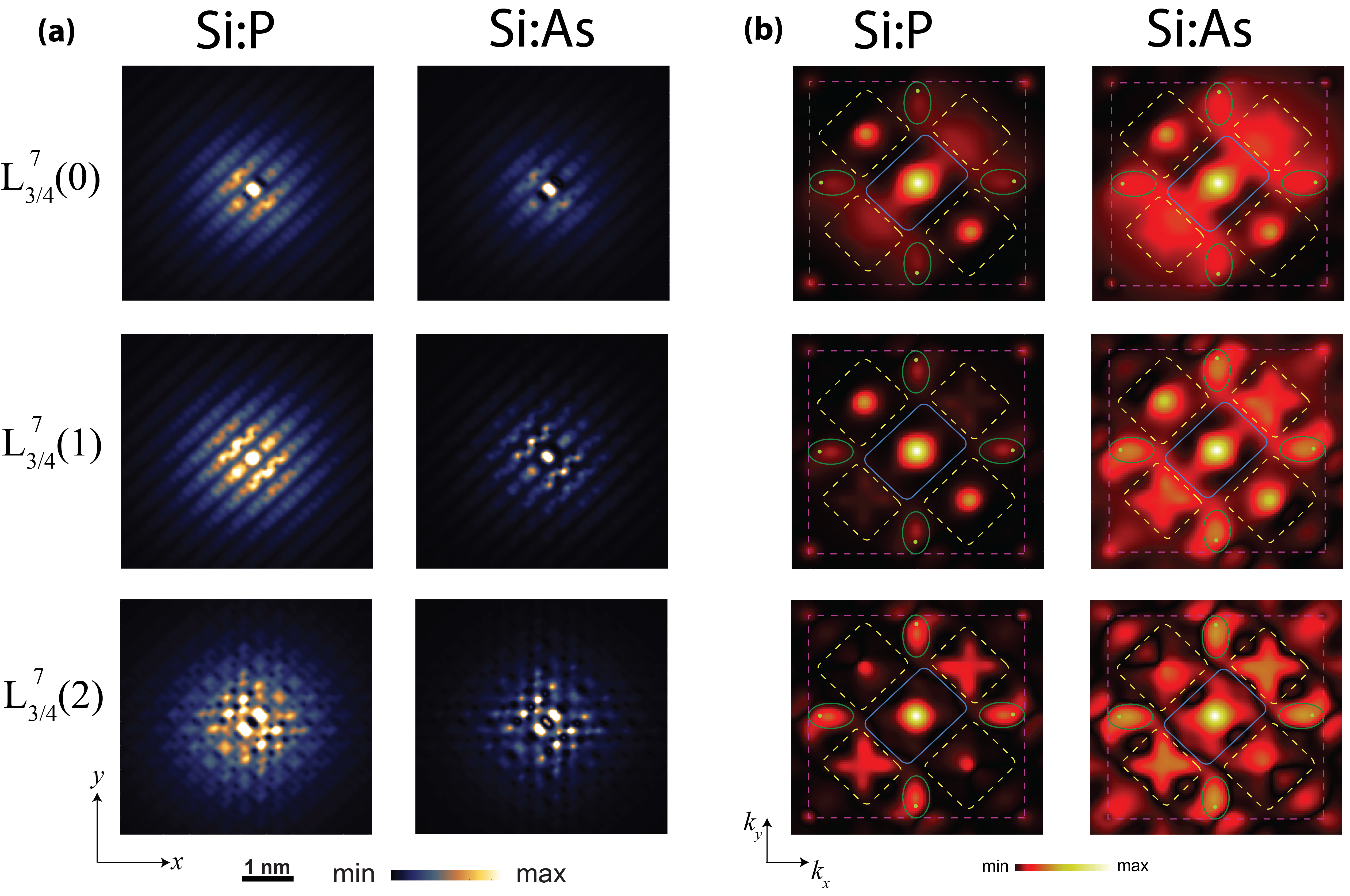}
\caption{(a) Real-space images of P and As donors in L$_{\rm 3/4}^{\rm 7}(\rm 0)$, L$_{\rm 3/4}^{\rm 7}(\rm 1)$, and L$_{\rm 3/4}^{\rm 7}(\rm 2)$ locations computed from the CCC$^{\rm NS}$ model. (b) Fourier transform spectrum of P and As donor images in L$_{\rm 3/4}^{\rm 7}(\rm 0)$, L$_{\rm 3/4}^{\rm 7}(\rm 1)$, and L$_{\rm 3/4}^{\rm 7}(\rm 2)$ positions computed from the CCC$^{\rm NS}$ model. In each image, the corners of the outer dashed purple box are reciprocal lattice vectors (2$\pi$/$a_0$) $(p,q)$, with $p$=$\pm$1 and $q$=$\pm$1. The ellipsoidal structures corresponding to valleys are found within the green ovals and the green dots indicate the position of the conduction band minima: $k_x$ = 0.82(2$\pi$/$a_0$) ($\pm1,0)$ and $k_y$ = 0.82(2$\pi$/$a_0$) ($0,\pm1)$. The region marked with blue boundary indicates probability envelope and the yellow dashed regions highlight the 2$\times$1 reconstruction-induced features. }
\label{fig:Fig4}
\end{figure*}


\newpage
\clearpage

\begin{thebibliography}{50}%
\makeatletter
\providecommand \@ifxundefined [1]{%
 \@ifx{#1\undefined}
}%
\providecommand \@ifnum [1]{%
 \ifnum #1\expandafter \@firstoftwo
 \else \expandafter \@secondoftwo
 \fi
}%
\providecommand \@ifx [1]{%
 \ifx #1\expandafter \@firstoftwo
 \else \expandafter \@secondoftwo
 \fi
}%
\providecommand \natexlab [1]{#1}%
\providecommand \enquote  [1]{``#1''}%
\providecommand \bibnamefont  [1]{#1}%
\providecommand \bibfnamefont [1]{#1}%
\providecommand \citenamefont [1]{#1}%
\providecommand \href@noop [0]{\@secondoftwo}%
\providecommand \href [0]{\begingroup \@sanitize@url \@href}%
\providecommand \@href[1]{\@@startlink{#1}\@@href}%
\providecommand \@@href[1]{\endgroup#1\@@endlink}%
\providecommand \@sanitize@url [0]{\catcode `\\12\catcode `\$12\catcode
  `\&12\catcode `\#12\catcode `\^12\catcode `\_12\catcode `\%12\relax}%
\providecommand \@@startlink[1]{}%
\providecommand \@@endlink[0]{}%
\providecommand \url  [0]{\begingroup\@sanitize@url \@url }%
\providecommand \@url [1]{\endgroup\@href {#1}{\urlprefix }}%
\providecommand \urlprefix  [0]{URL }%
\providecommand \Eprint [0]{\href }%
\providecommand \doibase [0]{http://dx.doi.org/}%
\providecommand \selectlanguage [0]{\@gobble}%
\providecommand \bibinfo  [0]{\@secondoftwo}%
\providecommand \bibfield  [0]{\@secondoftwo}%
\providecommand \translation [1]{[#1]}%
\providecommand \BibitemOpen [0]{}%
\providecommand \bibitemStop [0]{}%
\providecommand \bibitemNoStop [0]{.\EOS\space}%
\providecommand \EOS [0]{\spacefactor3000\relax}%
\providecommand \BibitemShut  [1]{\csname bibitem#1\endcsname}%
\let\auto@bib@innerbib\@empty
%</preamble>
\bibitem [{\citenamefont {Kane}(1998)}]{Kane_Nature_1998}%
  \BibitemOpen
  \bibfield  {author} {\bibinfo {author} {\bibfnamefont {B.~E.}\ \bibnamefont
  {Kane}},\ }\href@noop {} {\bibfield  {journal} {\bibinfo  {journal} {Nature}\
  }\textbf {\bibinfo {volume} {393}},\ \bibinfo {pages} {133} (\bibinfo {year}
  {1998})}\BibitemShut {NoStop}%
\bibitem [{\citenamefont {Hollenberg}\ \emph {et~al.}(2006)\citenamefont
  {Hollenberg}, \citenamefont {Greentree}, \citenamefont {Fowler},\ and\
  \citenamefont {Wellard}}]{Hollenberg_PRB_2006}%
  \BibitemOpen
  \bibfield  {author} {\bibinfo {author} {\bibfnamefont {L.}~\bibnamefont
  {Hollenberg}}, \bibinfo {author} {\bibfnamefont {A.~D.}\ \bibnamefont
  {Greentree}}, \bibinfo {author} {\bibfnamefont {A.~G.}\ \bibnamefont
  {Fowler}}, \ and\ \bibinfo {author} {\bibfnamefont {C.~J.}\ \bibnamefont
  {Wellard}},\ }\href@noop {} {\bibfield  {journal} {\bibinfo  {journal} {Phys.
  Rev. B}\ }\textbf {\bibinfo {volume} {74}},\ \bibinfo {pages} {045311}
  (\bibinfo {year} {2006})}\BibitemShut {NoStop}%
\bibitem [{\citenamefont {Hill}\ \emph {et~al.}(2015)\citenamefont {Hill},
  \citenamefont {Peretz}, \citenamefont {Hile}, \citenamefont {House},
  \citenamefont {Fuechsle}, \citenamefont {Rogge}, \citenamefont {Simmons},\
  and\ \citenamefont {Hollenberg}}]{Hill_science_2015}%
  \BibitemOpen
  \bibfield  {author} {\bibinfo {author} {\bibfnamefont {C.}~\bibnamefont
  {Hill}}, \bibinfo {author} {\bibfnamefont {E.}~\bibnamefont {Peretz}},
  \bibinfo {author} {\bibfnamefont {S.}~\bibnamefont {Hile}}, \bibinfo {author}
  {\bibfnamefont {M.}~\bibnamefont {House}}, \bibinfo {author} {\bibfnamefont
  {M.}~\bibnamefont {Fuechsle}}, \bibinfo {author} {\bibfnamefont
  {S.}~\bibnamefont {Rogge}}, \bibinfo {author} {\bibfnamefont {M.~Y.}\
  \bibnamefont {Simmons}}, \ and\ \bibinfo {author} {\bibfnamefont
  {L.}~\bibnamefont {Hollenberg}},\ }\href@noop {} {\bibfield  {journal}
  {\bibinfo  {journal} {Science Advances}\ }\textbf {\bibinfo {volume} {1}},\
  \bibinfo {pages} {e1500707} (\bibinfo {year} {2015})}\BibitemShut {NoStop}%
\bibitem [{\citenamefont {Pica}\ \emph {et~al.}(2016)\citenamefont {Pica},
  \citenamefont {Lovett}, \citenamefont {Bhatt}, \citenamefont {Schenkel},\
  and\ \citenamefont {Lyon}}]{Pica_PRB_2016}%
  \BibitemOpen
  \bibfield  {author} {\bibinfo {author} {\bibfnamefont {G.}~\bibnamefont
  {Pica}}, \bibinfo {author} {\bibfnamefont {B.~W.}\ \bibnamefont {Lovett}},
  \bibinfo {author} {\bibfnamefont {R.~N.}\ \bibnamefont {Bhatt}}, \bibinfo
  {author} {\bibfnamefont {T.}~\bibnamefont {Schenkel}}, \ and\ \bibinfo
  {author} {\bibfnamefont {S.~A.}\ \bibnamefont {Lyon}},\ }\href@noop {}
  {\bibfield  {journal} {\bibinfo  {journal} {Phys. Rev. B}\ }\textbf {\bibinfo
  {volume} {93}},\ \bibinfo {pages} {035306} (\bibinfo {year}
  {2016})}\BibitemShut {NoStop}%
\bibitem [{\citenamefont {Fuechsle}\ \emph {et~al.}(2012)\citenamefont
  {Fuechsle}, \citenamefont {Miwa}, \citenamefont {Mahapatra}, \citenamefont
  {Ryu}, \citenamefont {Lee}, \citenamefont {Warschkow}, \citenamefont
  {Hollenberg}, \citenamefont {Klimeck},\ and\ \citenamefont
  {Simmons}}]{Fuechsle_NN_2012}%
  \BibitemOpen
  \bibfield  {author} {\bibinfo {author} {\bibfnamefont {M.}~\bibnamefont
  {Fuechsle}}, \bibinfo {author} {\bibfnamefont {J.~A.}\ \bibnamefont {Miwa}},
  \bibinfo {author} {\bibfnamefont {S.}~\bibnamefont {Mahapatra}}, \bibinfo
  {author} {\bibfnamefont {H.}~\bibnamefont {Ryu}}, \bibinfo {author}
  {\bibfnamefont {S.}~\bibnamefont {Lee}}, \bibinfo {author} {\bibfnamefont
  {O.}~\bibnamefont {Warschkow}}, \bibinfo {author} {\bibfnamefont {L.~C.~L.}\
  \bibnamefont {Hollenberg}}, \bibinfo {author} {\bibfnamefont
  {G.}~\bibnamefont {Klimeck}}, \ and\ \bibinfo {author} {\bibfnamefont
  {M.~Y.}\ \bibnamefont {Simmons}},\ }\href@noop {} {\bibfield  {journal}
  {\bibinfo  {journal} {Nature Nanotechnology}\ }\textbf {\bibinfo {volume}
  {7}},\ \bibinfo {pages} {242} (\bibinfo {year} {2012})}\BibitemShut {NoStop}%
\bibitem [{\citenamefont {Weber}\ \emph {et~al.}(2012)\citenamefont {Weber}
  \emph {et~al.}}]{Weber_Science_2012}%
  \BibitemOpen
  \bibfield  {author} {\bibinfo {author} {\bibfnamefont {B.}~\bibnamefont
  {Weber}} \emph {et~al.},\ }\href@noop {} {\bibfield  {journal} {\bibinfo
  {journal} {Science}\ }\textbf {\bibinfo {volume} {335}},\ \bibinfo {pages}
  {64} (\bibinfo {year} {2012})}\BibitemShut {NoStop}%
\bibitem [{\citenamefont {Rahman}\ \emph {et~al.}(2007)\citenamefont {Rahman},
  \citenamefont {Wellard}, \citenamefont {Bradbury}, \citenamefont {Prada},
  \citenamefont {Cole}, \citenamefont {Klimeck},\ and\ \citenamefont
  {Hollenberg}}]{Rahman_PRL_2007}%
  \BibitemOpen
  \bibfield  {author} {\bibinfo {author} {\bibfnamefont {R.}~\bibnamefont
  {Rahman}}, \bibinfo {author} {\bibfnamefont {C.~J.}\ \bibnamefont {Wellard}},
  \bibinfo {author} {\bibfnamefont {F.~R.}\ \bibnamefont {Bradbury}}, \bibinfo
  {author} {\bibfnamefont {M.}~\bibnamefont {Prada}}, \bibinfo {author}
  {\bibfnamefont {J.~H.}\ \bibnamefont {Cole}}, \bibinfo {author}
  {\bibfnamefont {G.}~\bibnamefont {Klimeck}}, \ and\ \bibinfo {author}
  {\bibfnamefont {L.~C.~L.}\ \bibnamefont {Hollenberg}},\ }\href@noop {}
  {\bibfield  {journal} {\bibinfo  {journal} {Phys. Rev. Lett.}\ }\textbf
  {\bibinfo {volume} {99}},\ \bibinfo {pages} {036403} (\bibinfo {year}
  {2007})}\BibitemShut {NoStop}%
\bibitem [{\citenamefont {Usman}\ \emph
  {et~al.}(2015{\natexlab{a}})\citenamefont {Usman}, \citenamefont {Rahman},
  \citenamefont {Salfi}, \citenamefont {Bocquel}, \citenamefont {Voisin},
  \citenamefont {Rogge}, \citenamefont {Klimeck},\ and\ \citenamefont
  {Hollenberg}}]{Usman_JPCM_2015}%
  \BibitemOpen
  \bibfield  {author} {\bibinfo {author} {\bibfnamefont {M.}~\bibnamefont
  {Usman}}, \bibinfo {author} {\bibfnamefont {R.}~\bibnamefont {Rahman}},
  \bibinfo {author} {\bibfnamefont {J.}~\bibnamefont {Salfi}}, \bibinfo
  {author} {\bibfnamefont {J.}~\bibnamefont {Bocquel}}, \bibinfo {author}
  {\bibfnamefont {B.}~\bibnamefont {Voisin}}, \bibinfo {author} {\bibfnamefont
  {S.}~\bibnamefont {Rogge}}, \bibinfo {author} {\bibfnamefont
  {G.}~\bibnamefont {Klimeck}}, \ and\ \bibinfo {author} {\bibfnamefont
  {L.~C.~L.}\ \bibnamefont {Hollenberg}},\ }\href@noop {} {\bibfield  {journal}
  {\bibinfo  {journal} {J. Phys.: Cond. Matt.}\ }\textbf {\bibinfo {volume}
  {27}},\ \bibinfo {pages} {154207} (\bibinfo {year}
  {2015}{\natexlab{a}})}\BibitemShut {NoStop}%
\bibitem [{\citenamefont {Usman}\ \emph
  {et~al.}(2015{\natexlab{b}})\citenamefont {Usman}, \citenamefont {Hill},
  \citenamefont {Rahman}, \citenamefont {Klimeck}, \citenamefont {Simmons},
  \citenamefont {Rogge},\ and\ \citenamefont {Hollenberg}}]{Usman_PRB_2015}%
  \BibitemOpen
  \bibfield  {author} {\bibinfo {author} {\bibfnamefont {M.}~\bibnamefont
  {Usman}}, \bibinfo {author} {\bibfnamefont {C.~D.}\ \bibnamefont {Hill}},
  \bibinfo {author} {\bibfnamefont {R.}~\bibnamefont {Rahman}}, \bibinfo
  {author} {\bibfnamefont {G.}~\bibnamefont {Klimeck}}, \bibinfo {author}
  {\bibfnamefont {M.~Y.}\ \bibnamefont {Simmons}}, \bibinfo {author}
  {\bibfnamefont {S.}~\bibnamefont {Rogge}}, \ and\ \bibinfo {author}
  {\bibfnamefont {L.~C.~L.}\ \bibnamefont {Hollenberg}},\ }\href@noop {}
  {\bibfield  {journal} {\bibinfo  {journal} {Phys. Rev. B}\ }\textbf {\bibinfo
  {volume} {91}},\ \bibinfo {pages} {245209} (\bibinfo {year}
  {2015}{\natexlab{b}})}\BibitemShut {NoStop}%
\bibitem [{\citenamefont {Kalra}\ \emph {et~al.}(2014)\citenamefont {Kalra}
  \emph {et~al.}}]{Kalra_PRX_2014}%
  \BibitemOpen
  \bibfield  {author} {\bibinfo {author} {\bibfnamefont {R.}~\bibnamefont
  {Kalra}} \emph {et~al.},\ }\href@noop {} {\bibfield  {journal} {\bibinfo
  {journal} {Phys. Rev. X}\ }\textbf {\bibinfo {volume} {4}},\ \bibinfo {pages}
  {021044} (\bibinfo {year} {2014})}\BibitemShut {NoStop}%
\bibitem [{\citenamefont {Zwanenburg}\ \emph {et~al.}(2013)\citenamefont
  {Zwanenburg}, \citenamefont {Dzurak}, \citenamefont {Morello}, \citenamefont
  {Simmons}, \citenamefont {Hollenberg}, \citenamefont {Klimeck}, \citenamefont
  {Rogge}, \citenamefont {Coppersmith},\ and\ \citenamefont
  {Eriksson}}]{Zwanenburg_RMP_2013}%
  \BibitemOpen
  \bibfield  {author} {\bibinfo {author} {\bibfnamefont {F.}~\bibnamefont
  {Zwanenburg}}, \bibinfo {author} {\bibfnamefont {A.~S.}\ \bibnamefont
  {Dzurak}}, \bibinfo {author} {\bibfnamefont {A.}~\bibnamefont {Morello}},
  \bibinfo {author} {\bibfnamefont {M.~Y.}\ \bibnamefont {Simmons}}, \bibinfo
  {author} {\bibfnamefont {L.~C.~L.}\ \bibnamefont {Hollenberg}}, \bibinfo
  {author} {\bibfnamefont {G.}~\bibnamefont {Klimeck}}, \bibinfo {author}
  {\bibfnamefont {S.}~\bibnamefont {Rogge}}, \bibinfo {author} {\bibfnamefont
  {S.~N.}\ \bibnamefont {Coppersmith}}, \ and\ \bibinfo {author} {\bibfnamefont
  {M.~A.}\ \bibnamefont {Eriksson}},\ }\href@noop {} {\bibfield  {journal}
  {\bibinfo  {journal} {Rev. Mod. Phys.}\ }\textbf {\bibinfo {volume} {85}},\
  \bibinfo {pages} {961} (\bibinfo {year} {2013})}\BibitemShut {NoStop}%
\bibitem [{\citenamefont {Kohn}\ and\ \citenamefont
  {Luttinger}(1955)}]{Kohn_PR_1955}%
  \BibitemOpen
  \bibfield  {author} {\bibinfo {author} {\bibfnamefont {W.}~\bibnamefont
  {Kohn}}\ and\ \bibinfo {author} {\bibfnamefont {J.~M.}\ \bibnamefont
  {Luttinger}},\ }\href@noop {} {\bibfield  {journal} {\bibinfo  {journal}
  {Phys. Rev.}\ }\textbf {\bibinfo {volume} {98}},\ \bibinfo {pages} {915}
  (\bibinfo {year} {1955})}\BibitemShut {NoStop}%
\bibitem [{\citenamefont {Wilson}\ and\ \citenamefont
  {Feher}(1961)}]{Wilson_PR_1961}%
  \BibitemOpen
  \bibfield  {author} {\bibinfo {author} {\bibfnamefont {D.~K.}\ \bibnamefont
  {Wilson}}\ and\ \bibinfo {author} {\bibfnamefont {G.}~\bibnamefont {Feher}},\
  }\href@noop {} {\bibfield  {journal} {\bibinfo  {journal} {Phys. Rev.}\
  }\textbf {\bibinfo {volume} {124}},\ \bibinfo {pages} {1068} (\bibinfo {year}
  {1961})}\BibitemShut {NoStop}%
\bibitem [{\citenamefont {Pantelides}\ and\ \citenamefont
  {Sah}(1974)}]{Pantelides_Sah_PRB_1974}%
  \BibitemOpen
  \bibfield  {author} {\bibinfo {author} {\bibfnamefont {S.~T.}\ \bibnamefont
  {Pantelides}}\ and\ \bibinfo {author} {\bibfnamefont {C.~T.}\ \bibnamefont
  {Sah}},\ }\href@noop {} {\bibfield  {journal} {\bibinfo  {journal} {Phys.
  Rev. B}\ }\textbf {\bibinfo {volume} {10}},\ \bibinfo {pages} {621} (\bibinfo
  {year} {1974})}\BibitemShut {NoStop}%
\bibitem [{\citenamefont {Martins}\ \emph {et~al.}(2004)\citenamefont
  {Martins}, \citenamefont {Capaz},\ and\ \citenamefont
  {Koiller}}]{Martins_PRB_2004}%
  \BibitemOpen
  \bibfield  {author} {\bibinfo {author} {\bibfnamefont {A.~S.}\ \bibnamefont
  {Martins}}, \bibinfo {author} {\bibfnamefont {R.~B.}\ \bibnamefont {Capaz}},
  \ and\ \bibinfo {author} {\bibfnamefont {B.}~\bibnamefont {Koiller}},\
  }\href@noop {} {\bibfield  {journal} {\bibinfo  {journal} {Phys. Rev. B}\
  }\textbf {\bibinfo {volume} {69}},\ \bibinfo {pages} {085320} (\bibinfo
  {year} {2004})}\BibitemShut {NoStop}%
\bibitem [{\citenamefont {Friesen}(2005)}]{Friesen_PRL_2005}%
  \BibitemOpen
  \bibfield  {author} {\bibinfo {author} {\bibfnamefont {M.}~\bibnamefont
  {Friesen}},\ }\href@noop {} {\bibfield  {journal} {\bibinfo  {journal} {Phys.
  Rev. Lett.}\ }\textbf {\bibinfo {volume} {94}},\ \bibinfo {pages} {186403}
  (\bibinfo {year} {2005})}\BibitemShut {NoStop}%
\bibitem [{\citenamefont {Pica}\ \emph
  {et~al.}(2014{\natexlab{a}})\citenamefont {Pica}, \citenamefont {Wolfowicz},
  \citenamefont {Urdampilleta}, \citenamefont {Thewalt}, \citenamefont
  {Riemann}, \citenamefont {Abrosimov}, \citenamefont {Becker}, \citenamefont
  {Pohl}, \citenamefont {Morton}, \citenamefont {Bhatt}, \citenamefont {Lyon},\
  and\ \citenamefont {Lovett}}]{Pica_PRB_2014}%
  \BibitemOpen
  \bibfield  {author} {\bibinfo {author} {\bibfnamefont {G.}~\bibnamefont
  {Pica}}, \bibinfo {author} {\bibfnamefont {G.}~\bibnamefont {Wolfowicz}},
  \bibinfo {author} {\bibfnamefont {M.}~\bibnamefont {Urdampilleta}}, \bibinfo
  {author} {\bibfnamefont {M.~L.~W.}\ \bibnamefont {Thewalt}}, \bibinfo
  {author} {\bibfnamefont {H.}~\bibnamefont {Riemann}}, \bibinfo {author}
  {\bibfnamefont {N.~V.}\ \bibnamefont {Abrosimov}}, \bibinfo {author}
  {\bibfnamefont {P.}~\bibnamefont {Becker}}, \bibinfo {author} {\bibfnamefont
  {H.-J.}\ \bibnamefont {Pohl}}, \bibinfo {author} {\bibfnamefont {J.~J.~L.}\
  \bibnamefont {Morton}}, \bibinfo {author} {\bibfnamefont {R.~N.}\
  \bibnamefont {Bhatt}}, \bibinfo {author} {\bibfnamefont {S.~A.}\ \bibnamefont
  {Lyon}}, \ and\ \bibinfo {author} {\bibfnamefont {B.~W.}\ \bibnamefont
  {Lovett}},\ }\href@noop {} {\bibfield  {journal} {\bibinfo  {journal} {Phys.
  Rev. B}\ }\textbf {\bibinfo {volume} {90}},\ \bibinfo {pages} {195204}
  (\bibinfo {year} {2014}{\natexlab{a}})}\BibitemShut {NoStop}%
\bibitem [{\citenamefont {Gamble}\ \emph {et~al.}(2015)\citenamefont {Gamble}
  \emph {et~al.}}]{King_1}%
  \BibitemOpen
  \bibfield  {author} {\bibinfo {author} {\bibfnamefont {J.}~\bibnamefont
  {Gamble}} \emph {et~al.},\ }\href@noop {} {\bibfield  {journal} {\bibinfo
  {journal} {Phys. Rev. B}\ }\textbf {\bibinfo {volume} {91}},\ \bibinfo
  {pages} {235318} (\bibinfo {year} {2015})}\BibitemShut {NoStop}%
\bibitem [{\citenamefont {Wellard}\ and\ \citenamefont
  {Hollenberg}(2005)}]{Wellard_Hollenberg_PRB_2005}%
  \BibitemOpen
  \bibfield  {author} {\bibinfo {author} {\bibfnamefont {C.~J.}\ \bibnamefont
  {Wellard}}\ and\ \bibinfo {author} {\bibfnamefont {L.~C.~L.}\ \bibnamefont
  {Hollenberg}},\ }\href@noop {} {\bibfield  {journal} {\bibinfo  {journal}
  {Phys. Rev. B}\ }\textbf {\bibinfo {volume} {72}},\ \bibinfo {pages} {085202}
  (\bibinfo {year} {2005})}\BibitemShut {NoStop}%
\bibitem [{\citenamefont {Salfi}\ \emph {et~al.}(2014)\citenamefont {Salfi},
  \citenamefont {Mol}, \citenamefont {Rahman}, \citenamefont {Klimeck},
  \citenamefont {Simmons}, \citenamefont {Hollenberg},\ and\ \citenamefont
  {Rogge}}]{Salfi_NatMat_2014}%
  \BibitemOpen
  \bibfield  {author} {\bibinfo {author} {\bibfnamefont {J.}~\bibnamefont
  {Salfi}}, \bibinfo {author} {\bibfnamefont {J.~A.}\ \bibnamefont {Mol}},
  \bibinfo {author} {\bibfnamefont {R.}~\bibnamefont {Rahman}}, \bibinfo
  {author} {\bibfnamefont {G.}~\bibnamefont {Klimeck}}, \bibinfo {author}
  {\bibfnamefont {M.~Y.}\ \bibnamefont {Simmons}}, \bibinfo {author}
  {\bibfnamefont {L.~C.~L.}\ \bibnamefont {Hollenberg}}, \ and\ \bibinfo
  {author} {\bibfnamefont {S.}~\bibnamefont {Rogge}},\ }\href@noop {}
  {\bibfield  {journal} {\bibinfo  {journal} {Nature Materials}\ }\textbf
  {\bibinfo {volume} {13}},\ \bibinfo {pages} {605} (\bibinfo {year}
  {2014})}\BibitemShut {NoStop}%
\bibitem [{\citenamefont {Usman}\ \emph {et~al.}(2016)\citenamefont {Usman},
  \citenamefont {Bocquel}, \citenamefont {Salfi}, \citenamefont {Voisin},
  \citenamefont {Tankasala}, \citenamefont {Rahman}, \citenamefont {Simmons},
  \citenamefont {Rogge},\ and\ \citenamefont {Hollenberg}}]{Usman_NN_2016}%
  \BibitemOpen
  \bibfield  {author} {\bibinfo {author} {\bibfnamefont {M.}~\bibnamefont
  {Usman}}, \bibinfo {author} {\bibfnamefont {J.}~\bibnamefont {Bocquel}},
  \bibinfo {author} {\bibfnamefont {J.}~\bibnamefont {Salfi}}, \bibinfo
  {author} {\bibfnamefont {B.}~\bibnamefont {Voisin}}, \bibinfo {author}
  {\bibfnamefont {A.}~\bibnamefont {Tankasala}}, \bibinfo {author}
  {\bibfnamefont {R.}~\bibnamefont {Rahman}}, \bibinfo {author} {\bibfnamefont
  {M.~Y.}\ \bibnamefont {Simmons}}, \bibinfo {author} {\bibfnamefont
  {S.}~\bibnamefont {Rogge}}, \ and\ \bibinfo {author} {\bibfnamefont
  {L.}~\bibnamefont {Hollenberg}},\ }\href@noop {} {\bibfield  {journal}
  {\bibinfo  {journal} {Nature Nanotechnology}\ }\textbf {\bibinfo {volume}
  {11}},\ \bibinfo {pages} {763} (\bibinfo {year} {2016})}\BibitemShut
  {NoStop}%
\bibitem [{\citenamefont {Pica}\ \emph
  {et~al.}(2014{\natexlab{b}})\citenamefont {Pica} \emph
  {et~al.}}]{Pica2_PRB_2014}%
  \BibitemOpen
  \bibfield  {author} {\bibinfo {author} {\bibfnamefont {G.}~\bibnamefont
  {Pica}} \emph {et~al.},\ }\href@noop {} {\bibfield  {journal} {\bibinfo
  {journal} {Phys. Rev. B}\ }\textbf {\bibinfo {volume} {89}},\ \bibinfo
  {pages} {235306} (\bibinfo {year} {2014}{\natexlab{b}})}\BibitemShut
  {NoStop}%
\bibitem [{\citenamefont {Saraiva}\ \emph {et~al.}(2015)\citenamefont
  {Saraiva}, \citenamefont {Baena}, \citenamefont {Calderon},\ and\
  \citenamefont {Koiller}}]{Saraiva_arx_2014}%
  \BibitemOpen
  \bibfield  {author} {\bibinfo {author} {\bibfnamefont {A.~L.}\ \bibnamefont
  {Saraiva}}, \bibinfo {author} {\bibfnamefont {A.}~\bibnamefont {Baena}},
  \bibinfo {author} {\bibfnamefont {M.~J.}\ \bibnamefont {Calderon}}, \ and\
  \bibinfo {author} {\bibfnamefont {B.}~\bibnamefont {Koiller}},\ }\href@noop
  {} {\bibfield  {journal} {\bibinfo  {journal} {J. Phys.: Cond. Matt.}\
  }\textbf {\bibinfo {volume} {27}},\ \bibinfo {pages} {154208} (\bibinfo
  {year} {2015})}\BibitemShut {NoStop}%
\bibitem [{\citenamefont {Overhof}\ and\ \citenamefont
  {Gerstmann}(2004)}]{Overhof_PRL_2004}%
  \BibitemOpen
  \bibfield  {author} {\bibinfo {author} {\bibfnamefont {H.}~\bibnamefont
  {Overhof}}\ and\ \bibinfo {author} {\bibfnamefont {U.}~\bibnamefont
  {Gerstmann}},\ }\href@noop {} {\bibfield  {journal} {\bibinfo  {journal}
  {Phys. Rev. Lett.}\ }\textbf {\bibinfo {volume} {92}},\ \bibinfo {pages}
  {087602} (\bibinfo {year} {2004})}\BibitemShut {NoStop}%
\bibitem [{\citenamefont {Smith}\ \emph {et~al.}(2016)\citenamefont {Smith},
  \citenamefont {Budi}, \citenamefont {Per}, \citenamefont {Vogt},
  \citenamefont {Drumm}, \citenamefont {Hollenberg}, \citenamefont {Cole},\
  and\ \citenamefont {Russo}}]{Smith_arXiv_2016}%
  \BibitemOpen
  \bibfield  {author} {\bibinfo {author} {\bibfnamefont {J.~S.}\ \bibnamefont
  {Smith}}, \bibinfo {author} {\bibfnamefont {A.}~\bibnamefont {Budi}},
  \bibinfo {author} {\bibfnamefont {M.~C.}\ \bibnamefont {Per}}, \bibinfo
  {author} {\bibfnamefont {N.}~\bibnamefont {Vogt}}, \bibinfo {author}
  {\bibfnamefont {D.~W.}\ \bibnamefont {Drumm}}, \bibinfo {author}
  {\bibfnamefont {L.~C.~L.}\ \bibnamefont {Hollenberg}}, \bibinfo {author}
  {\bibfnamefont {J.~H.}\ \bibnamefont {Cole}}, \ and\ \bibinfo {author}
  {\bibfnamefont {S.~P.}\ \bibnamefont {Russo}},\ }\href@noop {} {\bibfield
  {journal} {\bibinfo  {journal} {arXiv:1612.00569}\ } (\bibinfo {year}
  {2016})}\BibitemShut {NoStop}%
\bibitem [{\citenamefont {Koiller}\ \emph {et~al.}(2002)\citenamefont
  {Koiller}, \citenamefont {Hu},\ and\ \citenamefont
  {Sarma}}]{Koiller_PRB_2002}%
  \BibitemOpen
  \bibfield  {author} {\bibinfo {author} {\bibfnamefont {B.}~\bibnamefont
  {Koiller}}, \bibinfo {author} {\bibfnamefont {X.}~\bibnamefont {Hu}}, \ and\
  \bibinfo {author} {\bibfnamefont {S.~D.}\ \bibnamefont {Sarma}},\ }\href@noop
  {} {\bibfield  {journal} {\bibinfo  {journal} {Phys. Rev. B}\ }\textbf
  {\bibinfo {volume} {66}},\ \bibinfo {pages} {115201} (\bibinfo {year}
  {2002})}\BibitemShut {NoStop}%
\bibitem [{\citenamefont {Ahmed}\ \emph {et~al.}(2009)\citenamefont {Ahmed},
  \citenamefont {Kharche}, \citenamefont {Rahman}, \citenamefont {Usman},
  \citenamefont {Lee}, \citenamefont {Ryu}, \citenamefont {Bae}, \citenamefont
  {Clark}, \citenamefont {Haley}, \citenamefont {Naumov}, \citenamefont
  {Saied}, \citenamefont {Korkusinski}, \citenamefont {Kennel}, \citenamefont
  {McLennan}, \citenamefont {Boykin},\ and\ \citenamefont
  {Klimeck}}]{Ahmed2009}%
  \BibitemOpen
  \bibfield  {author} {\bibinfo {author} {\bibfnamefont {S.}~\bibnamefont
  {Ahmed}}, \bibinfo {author} {\bibfnamefont {N.}~\bibnamefont {Kharche}},
  \bibinfo {author} {\bibfnamefont {R.}~\bibnamefont {Rahman}}, \bibinfo
  {author} {\bibfnamefont {M.}~\bibnamefont {Usman}}, \bibinfo {author}
  {\bibfnamefont {S.}~\bibnamefont {Lee}}, \bibinfo {author} {\bibfnamefont
  {H.}~\bibnamefont {Ryu}}, \bibinfo {author} {\bibfnamefont {H.}~\bibnamefont
  {Bae}}, \bibinfo {author} {\bibfnamefont {S.}~\bibnamefont {Clark}}, \bibinfo
  {author} {\bibfnamefont {B.}~\bibnamefont {Haley}}, \bibinfo {author}
  {\bibfnamefont {M.}~\bibnamefont {Naumov}}, \bibinfo {author} {\bibfnamefont
  {F.}~\bibnamefont {Saied}}, \bibinfo {author} {\bibfnamefont
  {M.}~\bibnamefont {Korkusinski}}, \bibinfo {author} {\bibfnamefont
  {R.}~\bibnamefont {Kennel}}, \bibinfo {author} {\bibfnamefont
  {M.}~\bibnamefont {McLennan}}, \bibinfo {author} {\bibfnamefont {T.~B.}\
  \bibnamefont {Boykin}}, \ and\ \bibinfo {author} {\bibfnamefont
  {G.}~\bibnamefont {Klimeck}},\ }\enquote {\bibinfo {title} {Multimillion atom
  simulations with nemo3d},}\ in\ \href {\doibase
  10.1007/978-0-387-30440-3_343} {\emph {\bibinfo {booktitle} {Encyclopedia of
  Complexity and Systems Science}}},\ \bibinfo {editor} {edited by\ \bibinfo
  {editor} {\bibfnamefont {R.~A.}\ \bibnamefont {Meyers}}}\ (\bibinfo
  {publisher} {Springer New York},\ \bibinfo {address} {New York, NY},\
  \bibinfo {year} {2009})\ pp.\ \bibinfo {pages} {5745--5783}\BibitemShut
  {NoStop}%
\bibitem [{\citenamefont {Nara}(1965)}]{Nara_JPSJ_1965}%
  \BibitemOpen
  \bibfield  {author} {\bibinfo {author} {\bibfnamefont {H.}~\bibnamefont
  {Nara}},\ }\href@noop {} {\bibfield  {journal} {\bibinfo  {journal} {J. Phys.
  Soc. Jap.}\ }\textbf {\bibinfo {volume} {20}},\ \bibinfo {pages} {778}
  (\bibinfo {year} {1965})}\BibitemShut {NoStop}%
\bibitem [{\citenamefont {Klimeck}\ \emph
  {et~al.}(2007{\natexlab{a}})\citenamefont {Klimeck}, \citenamefont {Ahmed},
  \citenamefont {Kharche}, \citenamefont {Korkusinski}, \citenamefont {Usman},
  \citenamefont {Parada},\ and\ \citenamefont {Boykin}}]{Klimeck_2}%
  \BibitemOpen
  \bibfield  {author} {\bibinfo {author} {\bibfnamefont {G.}~\bibnamefont
  {Klimeck}}, \bibinfo {author} {\bibfnamefont {S.}~\bibnamefont {Ahmed}},
  \bibinfo {author} {\bibfnamefont {N.}~\bibnamefont {Kharche}}, \bibinfo
  {author} {\bibfnamefont {M.}~\bibnamefont {Korkusinski}}, \bibinfo {author}
  {\bibfnamefont {M.}~\bibnamefont {Usman}}, \bibinfo {author} {\bibfnamefont
  {M.}~\bibnamefont {Parada}}, \ and\ \bibinfo {author} {\bibfnamefont
  {T.}~\bibnamefont {Boykin}},\ }\href@noop {} {\bibfield  {journal} {\bibinfo
  {journal} {IEEE Trans. Elect. Dev.}\ }\textbf {\bibinfo {volume} {54}},\
  \bibinfo {pages} {2090} (\bibinfo {year} {2007}{\natexlab{a}})}\BibitemShut
  {NoStop}%
\bibitem [{\citenamefont {Boykin}\ \emph {et~al.}(2004)\citenamefont {Boykin},
  \citenamefont {Klimeck},\ and\ \citenamefont {Oyafuso}}]{Boykin_PRB_2004}%
  \BibitemOpen
  \bibfield  {author} {\bibinfo {author} {\bibfnamefont {T.~B.}\ \bibnamefont
  {Boykin}}, \bibinfo {author} {\bibfnamefont {G.}~\bibnamefont {Klimeck}}, \
  and\ \bibinfo {author} {\bibfnamefont {F.}~\bibnamefont {Oyafuso}},\
  }\href@noop {} {\bibfield  {journal} {\bibinfo  {journal} {Phys. Rev. B}\
  }\textbf {\bibinfo {volume} {69}},\ \bibinfo {pages} {115201} (\bibinfo
  {year} {2004})}\BibitemShut {NoStop}%
\bibitem [{\citenamefont {Ramdas}\ and\ \citenamefont
  {Rodriguez}(1981)}]{Ramdas_RPP_1981}%
  \BibitemOpen
  \bibfield  {author} {\bibinfo {author} {\bibfnamefont {A.~K.}\ \bibnamefont
  {Ramdas}}\ and\ \bibinfo {author} {\bibfnamefont {S.}~\bibnamefont
  {Rodriguez}},\ }\href@noop {} {\bibfield  {journal} {\bibinfo  {journal}
  {Rep. Prog. Phys.}\ }\textbf {\bibinfo {volume} {44}},\ \bibinfo {pages}
  {1297} (\bibinfo {year} {1981})}\BibitemShut {NoStop}%
\bibitem [{\citenamefont {Salfi}\ \emph {et~al.}(2017)\citenamefont {Salfi}
  \emph {et~al.}}]{Salfi_arXiv_2017}%
  \BibitemOpen
  \bibfield  {author} {\bibinfo {author} {\bibfnamefont {J.}~\bibnamefont
  {Salfi}} \emph {et~al.},\ }\href@noop {} {\bibfield  {journal} {\bibinfo
  {journal} {arXiv:1706.09261}\ } (\bibinfo {year} {2017})}\BibitemShut
  {NoStop}%
\bibitem [{\citenamefont {Slater}(1930)}]{Slater_PR_1930}%
  \BibitemOpen
  \bibfield  {author} {\bibinfo {author} {\bibfnamefont {J.~C.}\ \bibnamefont
  {Slater}},\ }\href@noop {} {\bibfield  {journal} {\bibinfo  {journal} {Phys.
  Rev.}\ }\textbf {\bibinfo {volume} {36}},\ \bibinfo {pages} {57} (\bibinfo
  {year} {1930})}\BibitemShut {NoStop}%
\bibitem [{\citenamefont {Bardeen}(1961)}]{Bardeen_PRL_1961}%
  \BibitemOpen
  \bibfield  {author} {\bibinfo {author} {\bibfnamefont {J.}~\bibnamefont
  {Bardeen}},\ }\href@noop {} {\bibfield  {journal} {\bibinfo  {journal} {Phys.
  Rev. Lett.}\ }\textbf {\bibinfo {volume} {6}},\ \bibinfo {pages} {57}
  (\bibinfo {year} {1961})}\BibitemShut {NoStop}%
\bibitem [{\citenamefont {Chen}(90 I)}]{Chen_PRB_1990}%
  \BibitemOpen
  \bibfield  {author} {\bibinfo {author} {\bibfnamefont {C.~J.}\ \bibnamefont
  {Chen}},\ }\href@noop {} {\bibfield  {journal} {\bibinfo  {journal} {Phys.
  Rev. B}\ }\textbf {\bibinfo {volume} {42}},\ \bibinfo {pages} {8841}
  (\bibinfo {year} {1990-I})}\BibitemShut {NoStop}%
\bibitem [{\citenamefont {Blanco}\ \emph {et~al.}(2006)\citenamefont {Blanco}
  \emph {et~al.}}]{Blanco_PSS_2006}%
  \BibitemOpen
  \bibfield  {author} {\bibinfo {author} {\bibfnamefont {J.~M.}\ \bibnamefont
  {Blanco}} \emph {et~al.},\ }\href@noop {} {\bibfield  {journal} {\bibinfo
  {journal} {Prog. Surf. Sci.}\ }\textbf {\bibinfo {volume} {81}},\ \bibinfo
  {pages} {403} (\bibinfo {year} {2006})}\BibitemShut {NoStop}%
\bibitem [{\citenamefont {Chaika}\ \emph {et~al.}(2010)\citenamefont {Chaika},
  \citenamefont {Nazin}, \citenamefont {Semenov}, \citenamefont {Bozhko},
  \citenamefont {Lubben}, \citenamefont {Krasnikov}, \citenamefont {Radican},\
  and\ \citenamefont {Shvets}}]{Chaika_EPL_2010}%
  \BibitemOpen
  \bibfield  {author} {\bibinfo {author} {\bibfnamefont {A.~N.}\ \bibnamefont
  {Chaika}}, \bibinfo {author} {\bibfnamefont {S.~S.}\ \bibnamefont {Nazin}},
  \bibinfo {author} {\bibfnamefont {V.~N.}\ \bibnamefont {Semenov}}, \bibinfo
  {author} {\bibfnamefont {S.~I.}\ \bibnamefont {Bozhko}}, \bibinfo {author}
  {\bibfnamefont {O.}~\bibnamefont {Lubben}}, \bibinfo {author} {\bibfnamefont
  {S.~A.}\ \bibnamefont {Krasnikov}}, \bibinfo {author} {\bibfnamefont
  {K.}~\bibnamefont {Radican}}, \ and\ \bibinfo {author} {\bibfnamefont
  {I.~V.}\ \bibnamefont {Shvets}},\ }\href@noop {} {\bibfield  {journal}
  {\bibinfo  {journal} {Euro Phys. Lett.}\ }\textbf {\bibinfo {volume} {92}},\
  \bibinfo {pages} {46003} (\bibinfo {year} {2010})}\BibitemShut {NoStop}%
\bibitem [{\citenamefont {Teobaldi}\ \emph {et~al.}(2012)\citenamefont
  {Teobaldi}, \citenamefont {Inami}, \citenamefont {Kanasaki}, \citenamefont
  {Tanimura},\ and\ \citenamefont {Shluger}}]{Teobaldi_PRB_2012}%
  \BibitemOpen
  \bibfield  {author} {\bibinfo {author} {\bibfnamefont {G.}~\bibnamefont
  {Teobaldi}}, \bibinfo {author} {\bibfnamefont {E.}~\bibnamefont {Inami}},
  \bibinfo {author} {\bibfnamefont {J.}~\bibnamefont {Kanasaki}}, \bibinfo
  {author} {\bibfnamefont {K.}~\bibnamefont {Tanimura}}, \ and\ \bibinfo
  {author} {\bibfnamefont {A.~L.}\ \bibnamefont {Shluger}},\ }\href@noop {}
  {\bibfield  {journal} {\bibinfo  {journal} {Phys. Rev. B}\ }\textbf {\bibinfo
  {volume} {85}},\ \bibinfo {pages} {085433} (\bibinfo {year}
  {2012})}\BibitemShut {NoStop}%
\bibitem [{\citenamefont {Feher}(1959)}]{Feher_PR_1959}%
  \BibitemOpen
  \bibfield  {author} {\bibinfo {author} {\bibfnamefont {G.}~\bibnamefont
  {Feher}},\ }\href@noop {} {\bibfield  {journal} {\bibinfo  {journal} {Phys.
  Rev.}\ }\textbf {\bibinfo {volume} {114}},\ \bibinfo {pages} {1219} (\bibinfo
  {year} {1959})}\BibitemShut {NoStop}%
\bibitem [{\citenamefont {Brazdova}\ \emph {et~al.}(2017)\citenamefont
  {Brazdova}, \citenamefont {Bowler}, \citenamefont {Sinthiptharakoon},
  \citenamefont {Studer}, \citenamefont {Rahnejat}, \citenamefont {Curson},
  \citenamefont {Schofield},\ and\ \citenamefont {Fisher}}]{Brazdova_PRB_2016}%
  \BibitemOpen
  \bibfield  {author} {\bibinfo {author} {\bibfnamefont {V.}~\bibnamefont
  {Brazdova}}, \bibinfo {author} {\bibfnamefont {D.}~\bibnamefont {Bowler}},
  \bibinfo {author} {\bibfnamefont {K.}~\bibnamefont {Sinthiptharakoon}},
  \bibinfo {author} {\bibfnamefont {P.}~\bibnamefont {Studer}}, \bibinfo
  {author} {\bibfnamefont {A.}~\bibnamefont {Rahnejat}}, \bibinfo {author}
  {\bibfnamefont {N.}~\bibnamefont {Curson}}, \bibinfo {author} {\bibfnamefont
  {S.}~\bibnamefont {Schofield}}, \ and\ \bibinfo {author} {\bibfnamefont
  {A.}~\bibnamefont {Fisher}},\ }\href@noop {} {\bibfield  {journal} {\bibinfo
  {journal} {Phys. Rev. B}\ }\textbf {\bibinfo {volume} {95}},\ \bibinfo
  {pages} {075408} (\bibinfo {year} {2017})}\BibitemShut {NoStop}%
\bibitem [{\citenamefont {Usman}\ \emph {et~al.}(2011)\citenamefont {Usman}
  \emph {et~al.}}]{Usman_PRB2_2011}%
  \BibitemOpen
  \bibfield  {author} {\bibinfo {author} {\bibfnamefont {M.}~\bibnamefont
  {Usman}} \emph {et~al.},\ }\href@noop {} {\bibfield  {journal} {\bibinfo
  {journal} {Phys. Rev. B}\ }\textbf {\bibinfo {volume} {84}},\ \bibinfo
  {pages} {115321} (\bibinfo {year} {2011})}\BibitemShut {NoStop}%
\bibitem [{\citenamefont {Usman}(2012)}]{Usman_PRB3_2012}%
  \BibitemOpen
  \bibfield  {author} {\bibinfo {author} {\bibfnamefont {M.}~\bibnamefont
  {Usman}},\ }\href@noop {} {\bibfield  {journal} {\bibinfo  {journal} {Phys.
  Rev. B}\ }\textbf {\bibinfo {volume} {86}},\ \bibinfo {pages} {155444}
  (\bibinfo {year} {2012})}\BibitemShut {NoStop}%
\bibitem [{\citenamefont {Usman}\ \emph {et~al.}(2012)\citenamefont {Usman}
  \emph {et~al.}}]{Usman_Nanotechnology_2012}%
  \BibitemOpen
  \bibfield  {author} {\bibinfo {author} {\bibfnamefont {M.}~\bibnamefont
  {Usman}} \emph {et~al.},\ }\href@noop {} {\bibfield  {journal} {\bibinfo
  {journal} {Nanotechnology}\ }\textbf {\bibinfo {volume} {23}},\ \bibinfo
  {pages} {165202} (\bibinfo {year} {2012})}\BibitemShut {NoStop}%
\bibitem [{\citenamefont {Lee}\ \emph {et~al.}(2004)\citenamefont {Lee} \emph
  {et~al.}}]{Lee_PRB_2004}%
  \BibitemOpen
  \bibfield  {author} {\bibinfo {author} {\bibfnamefont {S.}~\bibnamefont
  {Lee}} \emph {et~al.},\ }\href@noop {} {\bibfield  {journal} {\bibinfo
  {journal} {Phys. Rev. B}\ }\textbf {\bibinfo {volume} {69}},\ \bibinfo
  {pages} {045316} (\bibinfo {year} {2004})}\BibitemShut {NoStop}%
\bibitem [{\citenamefont {Klimeck}\ \emph
  {et~al.}(2007{\natexlab{b}})\citenamefont {Klimeck}, \citenamefont {Ahmed},
  \citenamefont {Bae}, \citenamefont {Kharche}, \citenamefont {Clark},
  \citenamefont {Haley}, \citenamefont {Lee}, \citenamefont {Naumov},
  \citenamefont {Ryu}, \citenamefont {Saied}, \citenamefont {.Prada},
  \citenamefont {Korkusinski}, \citenamefont {Boykin},\ and\ \citenamefont
  {Rahman}}]{Klimeck_1}%
  \BibitemOpen
  \bibfield  {author} {\bibinfo {author} {\bibfnamefont {G.}~\bibnamefont
  {Klimeck}}, \bibinfo {author} {\bibfnamefont {S.}~\bibnamefont {Ahmed}},
  \bibinfo {author} {\bibfnamefont {H.}~\bibnamefont {Bae}}, \bibinfo {author}
  {\bibfnamefont {N.}~\bibnamefont {Kharche}}, \bibinfo {author} {\bibfnamefont
  {S.}~\bibnamefont {Clark}}, \bibinfo {author} {\bibfnamefont
  {B.}~\bibnamefont {Haley}}, \bibinfo {author} {\bibfnamefont
  {S.}~\bibnamefont {Lee}}, \bibinfo {author} {\bibfnamefont {M.}~\bibnamefont
  {Naumov}}, \bibinfo {author} {\bibfnamefont {H.}~\bibnamefont {Ryu}},
  \bibinfo {author} {\bibfnamefont {F.}~\bibnamefont {Saied}}, \bibinfo
  {author} {\bibfnamefont {M.}~\bibnamefont {.Prada}}, \bibinfo {author}
  {\bibfnamefont {M.}~\bibnamefont {Korkusinski}}, \bibinfo {author}
  {\bibfnamefont {T.~B.}\ \bibnamefont {Boykin}}, \ and\ \bibinfo {author}
  {\bibfnamefont {R.}~\bibnamefont {Rahman}},\ }\href@noop {} {\bibfield
  {journal} {\bibinfo  {journal} {IEEE Trans. Elect. Dev.}\ }\textbf {\bibinfo
  {volume} {54}},\ \bibinfo {pages} {2079} (\bibinfo {year}
  {2007}{\natexlab{b}})}\BibitemShut {NoStop}%
\bibitem [{\citenamefont {Craig}\ and\ \citenamefont
  {Smith}(1990)}]{Craig_SS_1990}%
  \BibitemOpen
  \bibfield  {author} {\bibinfo {author} {\bibfnamefont {B.~I.}\ \bibnamefont
  {Craig}}\ and\ \bibinfo {author} {\bibfnamefont {P.~V.}\ \bibnamefont
  {Smith}},\ }\href@noop {} {\bibfield  {journal} {\bibinfo  {journal} {Surface
  Science}\ }\textbf {\bibinfo {volume} {226}},\ \bibinfo {pages} {L55}
  (\bibinfo {year} {1990})}\BibitemShut {NoStop}%
\bibitem [{\citenamefont {Saraiva}\ \emph {et~al.}(2014)\citenamefont {Saraiva}
  \emph {et~al.}}]{Saraiva_1}%
  \BibitemOpen
  \bibfield  {author} {\bibinfo {author} {\bibfnamefont {A.}~\bibnamefont
  {Saraiva}} \emph {et~al.},\ }\href@noop {} {\bibfield  {journal} {\bibinfo
  {journal} {arXiv:1407.8224v1}\ } (\bibinfo {year} {2014})}\BibitemShut
  {NoStop}%
\bibitem [{\citenamefont {Wellard}\ \emph {et~al.}(2003)\citenamefont
  {Wellard}, \citenamefont {Hollenberg}, \citenamefont {Parisoli},
  \citenamefont {Kettle}, \citenamefont {Goan}, \citenamefont {McIntosh},\ and\
  \citenamefont {Jamieson1}}]{Wellard_PRB_2003}%
  \BibitemOpen
  \bibfield  {author} {\bibinfo {author} {\bibfnamefont {C.~J.}\ \bibnamefont
  {Wellard}}, \bibinfo {author} {\bibfnamefont {L.~C.~L.}\ \bibnamefont
  {Hollenberg}}, \bibinfo {author} {\bibfnamefont {F.}~\bibnamefont
  {Parisoli}}, \bibinfo {author} {\bibfnamefont {L.~M.}\ \bibnamefont
  {Kettle}}, \bibinfo {author} {\bibfnamefont {H.-S.}\ \bibnamefont {Goan}},
  \bibinfo {author} {\bibfnamefont {J.~A.~L.}\ \bibnamefont {McIntosh}}, \ and\
  \bibinfo {author} {\bibfnamefont {D.~N.}\ \bibnamefont {Jamieson1}},\
  }\href@noop {} {\bibfield  {journal} {\bibinfo  {journal} {Phys. Rev. B}\
  }\textbf {\bibinfo {volume} {68}},\ \bibinfo {pages} {195209} (\bibinfo
  {year} {2003})}\BibitemShut {NoStop}%
\bibitem [{\citenamefont {Lee}\ \emph {et~al.}(2001)\citenamefont {Lee} \emph
  {et~al.}}]{Lee_PRB_2001}%
  \BibitemOpen
  \bibfield  {author} {\bibinfo {author} {\bibfnamefont {S.}~\bibnamefont
  {Lee}} \emph {et~al.},\ }\href@noop {} {\bibfield  {journal} {\bibinfo
  {journal} {Phys. Rev. B}\ }\textbf {\bibinfo {volume} {63}},\ \bibinfo
  {pages} {195318} (\bibinfo {year} {2001})}\BibitemShut {NoStop}%
\bibitem [{\citenamefont {Nielsen}\ \emph {et~al.}(2012)\citenamefont
  {Nielsen}, \citenamefont {Rahman},\ and\ \citenamefont
  {Muller}}]{Nielsen_JAP}%
  \BibitemOpen
  \bibfield  {author} {\bibinfo {author} {\bibfnamefont {E.}~\bibnamefont
  {Nielsen}}, \bibinfo {author} {\bibfnamefont {R.}~\bibnamefont {Rahman}}, \
  and\ \bibinfo {author} {\bibfnamefont {R.~P.}\ \bibnamefont {Muller}},\
  }\href@noop {} {\bibfield  {journal} {\bibinfo  {journal} {J. Appl. Phys.}\
  }\textbf {\bibinfo {volume} {112}},\ \bibinfo {pages} {114304} (\bibinfo
  {year} {2012})}\BibitemShut {NoStop}%
\end{thebibliography}
\end{document}